\documentclass[12pt,a4paper]{iopart}
\pdfoutput=1

\usepackage{iopams}
\usepackage{graphicx}
\usepackage[breaklinks=true,colorlinks=true,linkcolor=blue,urlcolor=blue,citecolor=blue]{hyperref}

\begin{document}

\title{Isotropic finite-difference discretization of stochastic conservation
laws preserving detailed balance}

\author{Mahan Raj Banerjee}
\address{Jawaharlal Nehru Centre for Advanced Scientific Research, Jakkur,
Bangalore 560064, India}
\ead{mahanraj@jncasr.ac.in}

\author{Sauro Succi}
\address{Istituto Applicazioni Calcolo, CNR Roma via dei Taurini 9, 00185
Roma, Italy}
\ead{succi@iac.cnr.it}

\author{Santosh Ansumali\footnote{Corresponding author.}}
\address{Jawaharlal Nehru Centre for Advanced Scientific Research, Jakkur,
Bangalore 560064, India}
\ead{ansumali@jncasr.ac.in}

\author{R. Adhikari}
\address{The Institute of Mathematical Sciences-HBNI, CIT Campus, Taramani,
Chennai 600113,India}
\address{DAMTP, Centre for Mathematical Sciences, University of Cambridge, Wilberforce Road, Cambridge CB3 0WA, UK}
\ead{rjoy@imsc.res.in}

\begin{abstract}
The dynamics of thermally fluctuating conserved order parameters are
described by stochastic conservation laws. Thermal equilibrium in
such systems requires the dissipative and stochastic components of
the flux to be related by detailed balance. Preserving this relation
in spatial and temporal discretization is necessary to obtain solutions
that have fidelity to the continuum. Here, we propose a finite-difference
discretization that preserves detailed balance on the lattice, has
spatial error that is isotropic to leading order in lattice spacing,
and can be integrated accurately in time using a delayed difference
method. We benchmark the method for model B dynamics with a $\phi^{4}$
Landau free energy and obtain excellent agreement with analytical
results.
\end{abstract}
\maketitle

\section{Introduction}

Thermal fluctuations are an essential part of complex phenomena as
diverse as Brownian motion in colloidal suspensions\cite{batchelor,segre},
concentration fluctuations in semi-dilute polymer solutions \cite{helfand,milner,wu},
capillary waves at fluctuating interfaces\cite{privman,evans,sides},
critical dynamics in binary mixtures \cite{debye1,debye2,wignall,tanaka}
and pattern formation \cite{cross,mishra}. In these systems, stochastic
partial differential equations in the form of conservation laws are
used to describe the time evolution of conserved densities. Examples
include the fluctuating Cahn-Hilliard-Cook equation \cite{cahn,cook,shinozaki},
the fluctuating Navier-Stokes equations of Landau and Lifshitz\cite{landau,hohenberg},
the fluctuating lubrication equation \cite{stone2001,stone2005} and
models of electrohydrodynamic instabilities in electrospinning experiments
\cite{electrojet}.

The flux, in the conservation laws, which are coarse-grained expressions
of microscopically reversible Hamiltonian dynamics, must satisfy an
important constraint: the stochastic part of the flux cannot be chosen
independently but must be related through a detailed balance condition
to the irreversible part of the flux \cite{zwanzig2001nonequilibrium}.
This relation is necessary to ensure that the coarse-grained dynamics
yields a stationary state with a Gibbs distribution, or in other words,
that the dynamics expressed by the conservation law is consistent
with micro-reversibility.

Discretization in space \emph{and} time is a necessary step in seeking
numerical solutions to stochastic conservation laws. The discretization
of the \emph{spatial} part of the conservation law commonly requires
discrete analogs of the vector differential operators - the gradient,
divergence, and curl - and of the Laplacian. It is well-known that
the discrete differential operators do not always inherit all properties
of the continuum operators. In particular, special care is needed
to preserve the continuum ``div-grad-curl'' identities like the vanishing
of the discrete curl of a discrete gradient or of the discrete divergence
of a discrete curl. The discretization of the stochastic flux is more
involved compared to Langevin equations as it is the \emph{divergence}
of a Gaussian random field. For detailed balance in a stochastic conservation
law, it is necessary to ensure that the discrete divergence of a discrete
gradient is identical to the discrete Laplacian (see below). Common
discrete representations of these operators, constructed in the setting
of deterministic conservation laws, typically do not satisfy this
last property. Their use in stochastic conservation laws results in
a violation of the detailed balance condition \cite{petschek1983computer,desai,ibanes2000dynamics}.

Even when the above ``mimetic'' property of the discrete spatial
derivative operators is ensured, detailed balance can be broken by
the temporal discretization: the probability distribution in the stationary
state may acquire a dependence on the time step (and, hence, on the
kinetic coefficients), when detailed balance would explicitly rule
out such a dependence. To the best of our knowledge, much work is needed on the \emph{temporal} discretization of the Cahn-Hilliard-Cook equation to yield a stationary
distribution that is independent of the temporal time step \cite{delong2013temporal} so as to minimize the violation of detailed balance at the discrete level, if not to eliminate
it altogether.

Here, we propose a finite-difference discretization of a stochastic
conservation law following a semi-discretization strategy and illustrate
our general results with the specific example of model B \cite{hohenberg}.
The stochastic partial differential equation is first discretized
in space to yield a set of coupled stochastic ordinary differential
equations. In this, we use discretizations of the gradient and divergence
that ensure isotropic truncation errors to leading order in lattice
spacing \cite{thampi-iso,rashmi}. We then \emph{define} the Laplacian to be
a composition of these discrete operators, so that the identity $\nabla^2=\boldsymbol{\nabla}\cdot\boldsymbol{\nabla}$
is satisfied by construction. The resulting Laplacian operator is
negative semi-definite with a trivial null space: the only eigenvector
with a zero eigenvalue is the constant. The support of this Laplacian,
though, is larger than those of comparable accuracy commonly used
in finite-difference methods.

The temporal discretization of the semi-discrete system is completed
using a novel delayed difference scheme. A distinct advantage of this
integration scheme, deriving from the trivial null space of the Laplacian,
is that it does not produce spurious checker-board modes in two- and
three-dimensional spaces. The isotropy and trivial null space of the
``mimetic'' discretization of the Laplacian, together with the delayed
time integrator, yields a numerical method that is stable, accurate
and efficient. Our numerical results are for two-point correlation
functions and order parameter distributions are in excellent agreement
with the well-known analytical results for model B.

To contextualize the contributions of this manuscript, we briefly
survey previous work on the topic. The necessity of a consistent discretization
of the vector differential operators for satisfying in model H was
first pointed out by one of the authors in an unpublished report.
Similar ideas were expressed in the subsequent work of Delgado et. al.\cite{delgado} and Garcia et. al.\cite{donev2}
on the finite-volume discretization of the compressible isothermal fluctuating hydrodynamics at nanoscale and
the same for stochastic conservation law obtained from a large-volume expansion of the chemical master equation
for reacting and diffusing species respectively.
Thampi et. al.\cite{thampi} presented both spectral and finite-volume discretizations
of the order parameter equation in model H, using fluctuating discrete
kinetic theory to describe momentum conservation. A systematic study
of finite-volume discretization schemes preserving detailed balance
for a variety of stochastic conservation laws whose evolution is generated
by Poisson and dissipation brackets has recently been initiated. Several
alternatives to explicit temporal integration has been explored by
Torre et. al.\cite{espanol}. The purpose of this (possibly incomplete) survey is to
emphasize that our work here focuses on the combination of finite-difference
spatial discretizations, which are both simple and popular, with delayed
temporal integrators which mitigate some of the drawbacks of using
finite-difference spatial discretizations.

The remainder of the paper is organized as follows. In section \ref{sec:Model-B}
we present model B of the Halperin-Hohenberg classification of dynamical
critical phenomena \cite{hohenberg} and list those analytical results
used later in our benchmarking. Section \ref{sec:-Spatial-discretization}
describes the different topological properties of the discrete operators
preserving the FDR at lattice level in connection with the non-interacting
order parameter dynamics of a system subjected to a single phase bulk
free energy potential. Section IV depicts the asynchronous time discretization
method which is crucial to achieve the required stability and accuracy
for the FDR preserving discrete operators. In section V, we investigate
the cases of the interacting order parameter dynamics for a system
which is subjected to both single and two phase equilibrium potentials.
In all these cases we find an excellent agreement between the present
method and the analytical and pseudo-spectral results. In section
VI, we conclude with the multidimensional generalization of the present
work and draw a comparison of our method to the well-known cell-dynamical
method of Oono and Puri\cite{oono1987,oono1988}, showing that isotropic
differences and delay-difference integrators provide an independent
formulation of going beyond the Oono-Puri method.

\section{Model B\label{sec:Model-B}}

Model B is a stochastic partial differential equation for a conserved
scalar order parameter field $\phi(\mathbf{x},t)$ whose dynamics
is driven by a competition between deterministic thermodynamic forces
and stochastic forces of thermal origin \cite{hohenberg,cahn,cook}.
The equation of motion is
\begin{equation}
\partial_{t}\phi=\boldsymbol{\nabla\cdot}\left(M\boldsymbol{\nabla}\frac{\delta{\cal F}}{\delta\phi}\right)+\boldsymbol{\nabla}\cdot\boldsymbol{\xi},\label{eq:modelB}
\end{equation}
where $M$ is the order parameter mobility and ${\cal F}$, the Landau
free energy, is a functional of the order parameter
\begin{equation}
 {\cal F}=\int d^{d}\mathbf{x}\left[f\left(\phi\right)+\frac{1}{2}K(\boldsymbol{\nabla}\phi(\mathbf{x},t))^2\right].\label{eq:landau-functional}
\end{equation}
The local part of the free energy density is here taken to be $f(\phi)=\frac{1}{2}A\phi^2+\frac{1}{4}B\phi^{4}$
, where $A$ can be either positive or negative but $B$ is always
positive. The positive coefficient $K$ in the non-local part is related
to the energy cost for gradients in the order parameter. The stochastic
flux $\boldsymbol{\xi}(\mathbf{x},t)$ is a zero-mean Gaussian random
field whose correlation is local in both space and time,
\begin{equation}
\left<\boldsymbol{\xi}({\bf x},t)\boldsymbol{\xi}({\bf x}^{\prime},t^{\prime})\right>=2k_{B}TM\thinspace\mathbf{I}\thinspace\delta(t-t^{\prime})\delta({\bf x}^{\prime}-{\bf x}),\label{eq:FDT}
\end{equation}
where $k_{B}$ is the Boltzmann constant, $T$ is the temperature
and $\mathbf{I}$ is the identity matrix in the space of Cartesian
indices. This fluctuation-dissipation relation for the stochastic
flux ensures that the stationary probability distribution is
\[
P[\phi(\mathbf{x},t)]=Z^{-1}\exp\left(-\beta\mathcal{F}\right),
\]
where $Z$ is the partition function.

Model B has a Gaussian fixed point, corresponding to the parameters
$A,K>0$ and $B=0$. It also has a non-trivial Wilson-Fisher fixed
point when $A=0$ and $B,K>0$. In addition, it allows for two-phase
coexistence between the phases $\phi=\pm1$ when $A<0$ and $B,K>0$.
The ``domain wall'' between these two phases is described by the
$\phi^{4}$ soliton. Correlation functions in all three cases can
be calculated in closed form.

Here, our principal interest is in the Gaussian fixed point. The free
energy is quadratic and the order parameter distribution, consequently,
is Gaussian. The two-point correlation determines all remaining correlation
functions. In Fourier space, it is

\begin{equation}
\langle\phi(\mathbf{q})\phi(-\mathbf{q}\mathbf{)}\rangle=\frac{k_{B}T}{A+Kq^2}.\label{eq:2pnt-corr}
\end{equation}
A first check on the accuracy of the discrete numerical method is
provided by a comparison with the two-point correlation function.
Away from the Gaussian fixed point, a more stringent check is provided
by a comparison with the distribution of the order parameter. Below,
we use both these checks to validate our numerical methods.

\section{Spatial discretization and detailed balance \label{sec:-Spatial-discretization}}

In this section we discretize Model B in space to show how naive discretizations
break detailed balance and how detailed balance can be restored by
a redefinition of the discrete Laplacian. This analysis most illustrative
without the additional complication of order parameter non-linearity
and therefore, we shall restrict ourselves to the Gaussian phase,
though the results obtained will be generally applicable. In the Gaussian
phase the equation of motion is linear and, for a constant mobility,
takes the form
\begin{equation}
\partial_{t}\phi(\mathbf{x},t)=M\nabla^2(A-K\nabla^2)\phi(\mathbf{x},t)+\boldsymbol{\nabla}\cdot\boldsymbol{\xi}(\mathbf{x},t).\label{CHC}
\end{equation}

Let us denote the discrete gradient, divergence and Laplacian by $\tilde{\boldsymbol{\nabla}}$,
$\tilde{\boldsymbol{\nabla\cdot}}$ and $\tilde{\nabla}^2$ respectively.
It follows that the equation of motion of the discretely sampled field
is
\begin{equation}
\partial_{t}\phi(\mathbf{x},t)=M\tilde{\nabla}^2(A-K\tilde{\nabla}^2)\phi(\mathbf{x},t)+\boldsymbol{\tilde{\nabla}}\cdot\boldsymbol{\xi}.
\end{equation}
It is then a straightforward exercise to show, using the fluctuation-dissipation
relation for the random flux, that the two-point correlation function
of the Fourier modes of the order parameter is given by

\begin{equation}
\langle\phi(\mathbf{q})\phi(-\mathbf{q}\mathbf{)}\rangle=\left(\frac{\tilde{\boldsymbol{\nabla}}_{\mathbf{q}}\cdot\tilde{\boldsymbol{\nabla}}_{\mathbf{q}}}{\tilde{\nabla}_{\mathbf{q}}^2}\right)\frac{k_{B}T}{A-K\tilde{\nabla}_{\mathbf{q}}^2},
\end{equation}
where the subscripts indicate the Fourier transforms of the respective
operators. Comparing with the two-point correlation function of the
continuum theory, Eq.\ref{eq:2pnt-corr}, it is evident that the
discrete two-point correlation function will contain the factor
\begin{equation}
\mathcal{R}(\mathbf{q)=\frac{\tilde{\boldsymbol{\nabla}}_{\mathbf{q}}\cdot\tilde{\boldsymbol{\nabla}}_{\mathbf{q}}}{\tilde{\nabla}_{\mathbf{q}}^2}}\label{Rfactor}
\end{equation}
which, generally, will differ from unity. To ensure this ``equilibrium
ratio'' to be unity for all wave numbers requires that the Fourier
transform of the discrete gradient, divergence and Laplacian operators
be related exactly as in the continuum.

To illustrate this analysis with a simple example, consider the standard
central-difference stencils in one dimension, for which
\begin{equation}
\tilde{\nabla}\phi(x) = \frac{\phi(x+\delta x)-\phi(x-\delta x)}{2\delta x},\label{cd2grad}
\end{equation}
\begin{equation}
\tilde{\nabla}^2\phi(x) = \frac{\phi(x+\delta x)-2\phi(x)+\phi(x-\delta x)}{(\delta x)^2}.\label{cd2lap}
\end{equation}
Using Fourier transform of Eq.\ref{cd2grad} and Eq.\ref{cd2lap} and using Eq.\ref{Rfactor} it can be shown that,
\begin{equation}
\mathcal{R}(\mathbf{q})=\frac{\cos(2q\delta x)-1}{4\cos(q\delta x)-4}.
\end{equation}
This expression tends to unity as $q$ tends to zero but is less than
unity for all other values of $q$ in the first Brillouin zone $|q|\leq\pi$.
Therefore, the use of such a discretization will, even for a (hypothetically)
perfect temporal integrator, introduce spurious wave number dependence
in the two-point correlation. This artifact of the standard discretization
has been noted earlier \cite{thampi}.

However, if the Laplacian is \emph{defined} as
\begin{equation}
\tilde{\nabla}^2\phi=\tilde{\nabla}\cdot\tilde{\nabla}\phi=\frac{\phi(x+2\delta x)-2\phi(x)+\phi(x-2\delta x)}{4(\delta x)^2},\label{cd2divgrad}
\end{equation}
repeating the above exercise shows that $\mathcal{R}(\mathbf{q})$
is unity for \emph{all} wave numbers and the semi-discretization ensures
that the correlation function approximates that of the continuum and
it is free of spurious wave number dependent contributions. The Fourier
transform of this Laplacian is
\begin{equation}
\tilde{\nabla}_{{\bf q}}^2=\frac{\cos(2q\delta x)-1}{2(\delta x)^2}=-\frac{\sin^2(q\delta x)}{(\delta x)^2}
\end{equation}
and is obviously negative semi-definite. The only null eigenvector
in the first Brillouin zone is a constant.

\section{Explicit time integrators}

The use of the Laplacian in Eq.\ref{cd2divgrad} in conventional
time discretization schemes will lead to reduced overall accuracy,
which can be seen by considering a simple but illustrative example
of the diffusion equation
\begin{equation}
\partial_{t}\phi=D\,\partial_{x}^2\phi\label{diffusion}
\end{equation}
for the scalar field $\phi$. Using central differences and forward
Euler for spatial and temporal discretization respectively, the resulting
difference equation is

\begin{equation}
\phi_{i}^{n+1}=\phi_{i}^{n}+\frac{\alpha}{(m+1)^2}\left[\phi_{i+m+1}^{n}+\phi_{i-m-1}^{n}-2\phi_{i}^{n}\right],\label{eulercd2}
\end{equation}
where $\phi_{i}^{n}\equiv\phi(i\Delta\mathbf{x},n\Delta t)$ and $\alpha=D\,\Delta t/\delta x^2$
is the Courant\textendash Friedrichs\textendash Lewy (CFL) number.
The Laplacians of Eq.\ref{cd2lap} and Eq.\ref{cd2divgrad} correspond
to $m=0$ and $m=1$ respectively. The numerical stability of the
difference scheme can be analyzed using the von Neumann method. The
amplification factor for Eq.\ref{eulercd2} is easily obtained to
be
\begin{equation}
\lambda=1-\frac{4\alpha}{(m+1)^2}\sin^2\left[\frac{(m+1)\,q\delta x}{2}\right]\label{eq:eulercd2fourier}
\end{equation}
and thus stability, $|\lambda|<1$, requires
\begin{equation}
0\leq\alpha\leq\frac{(m+1)^2}{2}.\label{eq:widerStencil}
\end{equation}

The continuum limit of Eq.\ref{eulercd2} under the ``diffusive
scaling'' $\delta x\sim O(\epsilon)$ and $\Delta t\sim O(\epsilon^2)$,
correct to $O(\epsilon^{4})$, is
\begin{equation}
\frac{\partial\phi}{\partial t}+\frac{\Delta t}{2}\frac{\partial^2\phi}{\partial t^2}=D\left(\frac{\partial^2\phi}{\partial x^2}+\frac{(m+1)^2\delta x^2}{12}\frac{\partial^{4}\phi}{\partial x^{4}}\right).\label{evolDFTCS-1}
\end{equation}
This equation can be simplified using using the Cauchy-Kowalewsky
backward error analysis, where all higher order time and mixed derivatives
are estimated by space derivatives obtained using the differential
equation itself. For example, by taking a derivative in time of the
evolution Eq.\ref{evolDFTCS-1}, we can estimate
\begin{equation}
\frac{\partial^2\phi}{\partial t^2}=D^2\,\frac{\partial^{4}\phi}{\partial x^{4}}+O(\epsilon^2).\label{tD3-1}
\end{equation}
Thus, the effective differential equation with an error of $O(\epsilon^{4})$
is

\begin{equation}
\frac{\partial\phi}{\partial t} =D\,\frac{\partial^2\phi}{\partial x^2}+\frac{D\,\delta x^2}{2}\frac{\partial^{4}\phi}{\partial x^{4}}\mathcal{\,I}^{{\rm CD2}}(m,\alpha),
\end{equation}
where the transport coefficient associated with the biharmonic operator
is
\[
\mathcal{I}^{{\rm CD2}}(m,\alpha)=\frac{(m+1)^2}{6}-\alpha.
\]
Based on this effective differential equation, we can analyze the
accuracy for different values of $m$ and thus the effect of a wider
stencil on accuracy. The trade-off in using the ``mimetic'' stencil
is now obvious: the wider stencil has a lower accuracy as $|\mathcal{I}^{{\rm CD2}}(1,\alpha)|>|\mathcal{I}^{{\rm CD2}}(0,\alpha)|$
in the parameter range $0\leq\alpha\leq1/2$ where the schemes are
stable.

\section{Delayed time integrators}

The above shortcoming of the combination of the spatial ``mimetic''
Laplacian and the temporal explicit integrator can be remedied by
using a recently introduced delayed-in-time integration scheme \cite{ansumali}.
This scheme is motivated by the observation that computing derivatives
on a wider stencil, while using spatial data from earlier times, can
dramatically improve both stability and accuracy \cite{ansumali}.

The delayed integrator applied to Eq.\ref{diffusion} gives
\begin{equation}
\phi_{i}^{n+1}=\phi_{i}^{n}+\frac{\alpha}{(m+1)^2}\left[\phi_{i+m+1}^{n-m}+\phi_{i-m-1}^{n-m}-2\phi_{i}^{n-m}\right],\label{deldiff}
\end{equation}
which should be compared with Eq.\ref{eulercd2} for $m=0$. The
amplification factor for this scheme obeys

\begin{equation}
\lambda^{m+1}-\lambda^{m}+\frac{4\alpha}{(m+1)^2}\sin^2\left(\frac{(m+1)\,q\delta x}{2}\right)=0,\label{baseVN-2}
\end{equation}
which can be solved for $m=1$ to obtain
\begin{equation}
\lambda=\frac{1}{2}\left(1\pm\sqrt{1-4\alpha\sin^2\left(q\delta x\right)}\right).\label{eq:roots}
\end{equation}
Hence, to satisfy $|\lambda|\leq1$, we have the stability condition
\[
\alpha\leq1,
\]
which implies a gain in stability compared to $m=0$. However, the
gain is less than naive use of wider stencil (Eq.\ref{eq:widerStencil}).
Here, we remind that wider stencil leads to better stability but much
lower accuracy. However, the delayed scheme removes this problem of
lower accuracy associated with wider stencil. This can be seen from
the effective differential equation corresponding to the delayed scheme.
In order to obtain such effective equation, similar to previous section,
we use Cauchy-Kowalewski procedure. We write Eq.\ref{deldiff} in
differential form using Taylor series as
\begin{equation}
\frac{\partial\phi}{\partial t}+\frac{\delta t}{2}\frac{\partial^2\phi}{\partial t^2}=D\left(\frac{\partial^2\phi}{\partial x^2}-m\delta t\frac{\partial^{3}\phi}{\partial t\partial x^2}+\frac{(m+1)^2\delta x^2}{12}\frac{\partial^{4}\phi}{\partial x^{4}}
\right)+O(\epsilon^{3})\label{evolDFTCS-2}
\end{equation}
and then replacing the time and the mixed derivatives using evolution
equation (Eq.\ref{evolDFTCS-2}), to obtain the effective differential
equation at the leading order as

\begin{equation}
\frac{\partial\phi}{\partial t}=D\,\frac{\partial^2\phi}{\partial x^2}+\frac{D\,\delta x^2}{2}\frac{\partial^{4}\phi}{\partial x^{4}}\mathcal{I}^{{\rm Delayed}}(m,\alpha)+O(\epsilon^{3}),
\label{eq:EDEDel}
\end{equation}
where,
\[
\mathcal{I}^{{\rm Delayed}}(m,\alpha)=\frac{(m+1)^2}{6}-\alpha(2m+1).
\]
 Thus, for $m=1$, Eq.\ref{eq:EDEDel} implies
\begin{equation}
\partial_{t}\phi=D\,\partial_{x}^2\phi+\frac{D\,\delta x^2}{2}\left[\frac{2}{3}-3\alpha\right]\partial_{x}^{4}\phi.\label{effdiffdel}
\end{equation}

\begin{figure}
\centering
\includegraphics[scale=0.35]{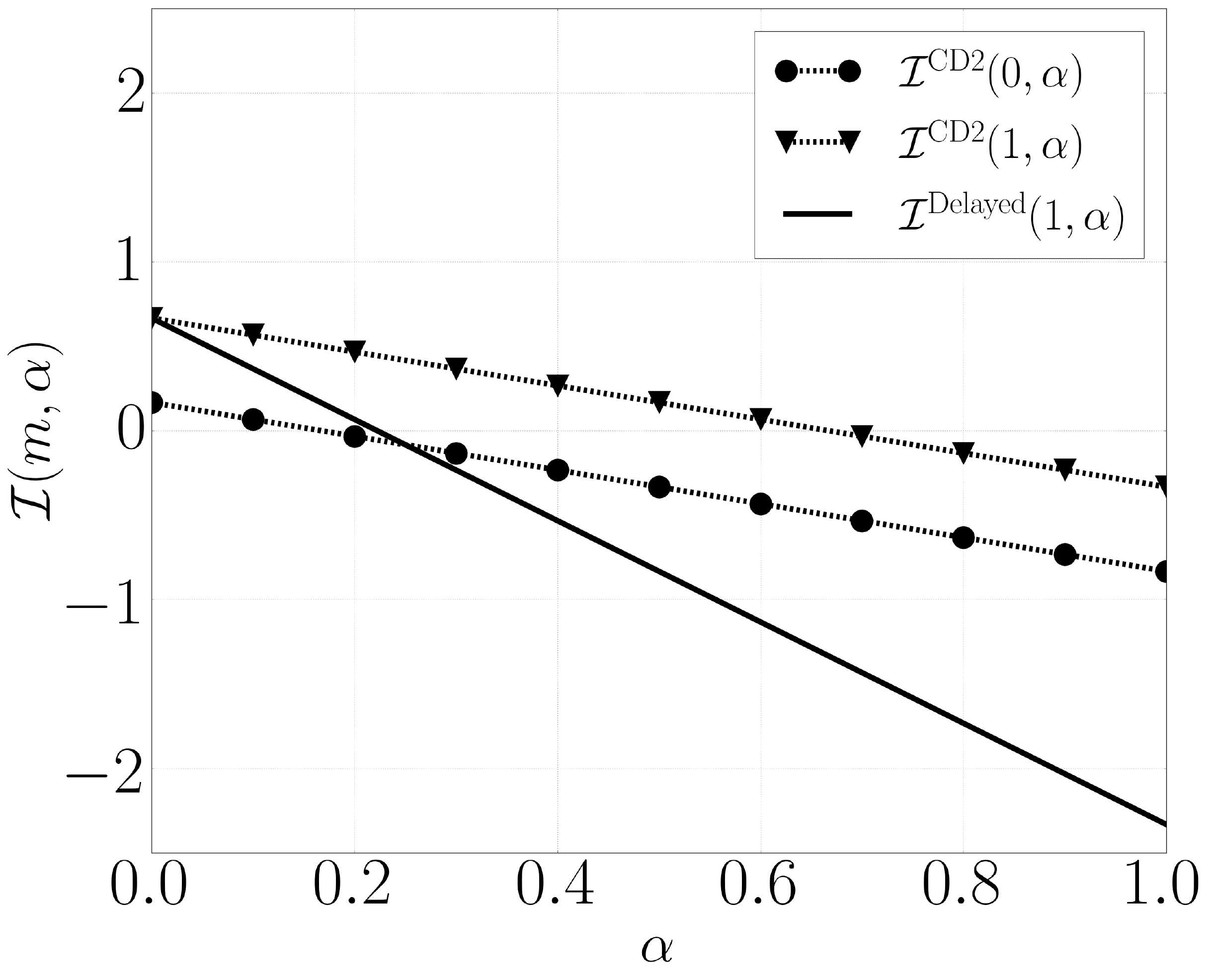}\caption{\label{accuracy}Variation of pre-factors $\mathcal{I}^{{\rm CD2}}(m)$
and $\mathcal{I}^{{\rm Delayed}}(m)$ of with CFL($\alpha$).}
\end{figure}

\begin{figure}
\centering
\includegraphics[scale=0.2]{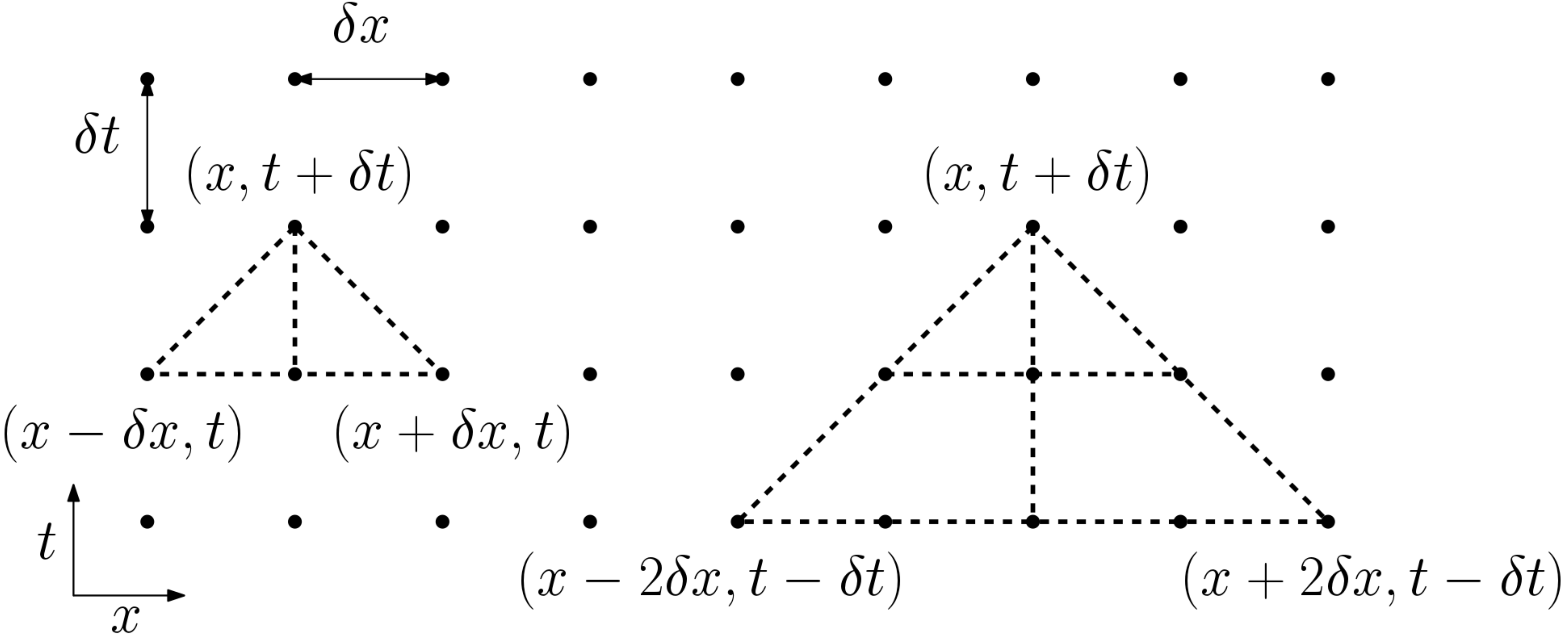}\caption{\label{Stencil}Stencil for Standard(left) and Delayed(right) schemes
and domain of dependence.}
\end{figure}

Thus, for $\alpha>0.25$ the delayed scheme has higher accuracy $\mathcal{I}^{{\rm Delayed}}(1,\alpha)$
than the standard CD2 schemes with $\mathcal{I}^{{\rm CD2}}(0,\alpha)$
as well as naive scheme with wider stencil $\mathcal{I}^{{\rm CD2}}(1,\alpha)$.
This is illustrated in Fig.\ref{accuracy}, where the pre-factors
$\mathcal{I}^{{\rm CD2}}(m,\alpha)$ and $\mathcal{I}^{{\rm Delayed}}(m,\alpha)$,
are plotted with respect to the CFL($\alpha$)

Thus, long time integration can be efficiently performed by the delayed
scheme, as it enhances the stability by a factor $2.0$ as compared
to that of the usual CD2 scheme, without reduction in accuracy. This
is evident from the schematics (Fig.\ref{Stencil}), which shows that
data taken from past requires wider stencil($2\delta x$) and thus,
widening of stencil compensates for error due to the time delay.

\section{Delayed time integrators in multi-dimensional space}

In last section, we have shown that the delayed scheme leads to better
stability and accuracy for discrtized diffusion equation in one dimension.
However, for the multi-dimensional extension of the delayed scheme,
\begin{equation}
\phi_{i}^{n+1}=\phi_{i}^{n}+D\,\delta t\tilde{\Delta}\phi_{i}^{n-1},\label{isodel}
\end{equation}
the increase in accuracy and stability requires further restrictions
on the form of discrete the operators. This can be seen by repeating
the analysis of previous section on multi-dimensional scheme. In this
case, similar to the previous section, we write Eq.\ref{deldiff}
in differential form using Taylor series as
\begin{equation}
\frac{\partial\phi}{\partial t}+\frac{\delta t}{2}\frac{\partial^2\phi}{\partial t^2}=D\left(\tilde{\Delta}\phi-\delta t\frac{\partial\tilde{\Delta}\phi}{\partial t}\right).\label{evolDFTCS-Multi}
\end{equation}
 If the discrete Laplacian preserves an isotropic structure at least
at the leading order, with $a$ as a stencil dependent constant, i.e,
\begin{equation}
\tilde{\Delta}=\nabla^2+a\delta x^2\nabla^2\nabla^2+\cdots,\label{isotropy}
\end{equation}
 the effective differential equation for the discrete analog of diffusion
equation can be written as

\begin{equation}
\partial_{t}\phi=D\,\nabla^2\left[1+\left(a\,\delta x^2-\frac{3D\,\Delta t}{2}\right)\nabla^2\right]\phi.\label{deleffode}
\end{equation}
 Similar to the 1-D case, this equation for the delayed scheme also
has better stability and accuracy. However, conventional discrete
operators such as central difference operators do not satisfy Eq.\ref{isotropy}\cite{thampi-iso,oono1987,oono1988}.
This can be seen by Taylor series expansion of central difference
Laplacian operator in 3-D

\begin{equation}
\tilde{\Delta}^{{\rm CD2}}=\nabla^2+\frac{\delta x^2}{12}(\partial_{x}^{4}+\partial_{y}^{4}+\partial_{z}^{4}).
\end{equation}
 Thus, we use recently proposed lattice differential operators\cite{thampi-iso,rashmi},
where discrete operators are constructed from lattice kinetic models.
In this approach, the basic discrete vector operators viz. Gradient($\tilde{\boldsymbol{\nabla}}$),
Divergence($\tilde{\boldsymbol{\nabla}}\cdot$) and Curl($\tilde{\boldsymbol{\nabla}}\wedge$)
for a given vector field $\mathbf{\Phi}$ on this lattice are formulated
as
\begin{eqnarray}
 & \tilde{\boldsymbol{\nabla}}^{{\rm iso}}\mathbf{\Phi}=\frac{1}{\delta x}\sum_{i=1}^{N}w_{i}\mathbf{\hat{c}_{i}}\mathbf{\Phi}\left(\mathbf{r}_{i}+\mathbf{c}_{i}\right),\label{LBgrad}\\
 & \tilde{\boldsymbol{\nabla}}^{{\rm iso}}\cdot\mathbf{\Phi}=\frac{1}{\delta x}\sum_{i=1}^{N}w_{i}\mathbf{\hat{c}_{i}}\cdot\mathbf{\Phi}\left(\mathbf{r}_{i}+\mathbf{c}_{i}\right),\label{LBdiv}\\
 & \tilde{\boldsymbol{\nabla}}^{{\rm iso}}\wedge\mathbf{\Phi}=\frac{1}{\delta x}\sum_{i=1}^{N}w_{i}\mathbf{\hat{c}_{i}}\wedge\mathbf{\Phi}\left(\mathbf{r}_{i}+\mathbf{c}_{i}\right),\label{LBcurl}
\end{eqnarray}
with the set of connecting vectors on stencil as $\mathbf{c}_{i}$
with $i=1,\cdots N$, where, $N$ being the total number of neighbours,
and the corresponding wights $w_{i}$ (normalized to one ($\sum_{i}^{N}w_{i}=1$)
are chosen so as to satisfy 
\begin{equation}
\sum_{i}^{N}w_{i}c_{i\alpha}c_{i\beta}=A\delta_{\alpha\beta},\;\sum_{i}^{N}w_{i}c_{i\alpha}c_{i\beta}c_{i\kappa}c_{i\eta}=B\Delta_{\alpha\beta\kappa\eta},
\end{equation}
with $\Delta_{\alpha\beta\kappa\eta}$ being the fourth order isotropic
Kronecker-delta and the connecting vectors are chosen such that $\sum_{i}^{N}w_{i}c_{i\alpha}=0$.
These conditions on weights ensure the isotropy of discrete operators
up to the leading orders.

\begin{figure}
\centering \includegraphics[scale=0.5]{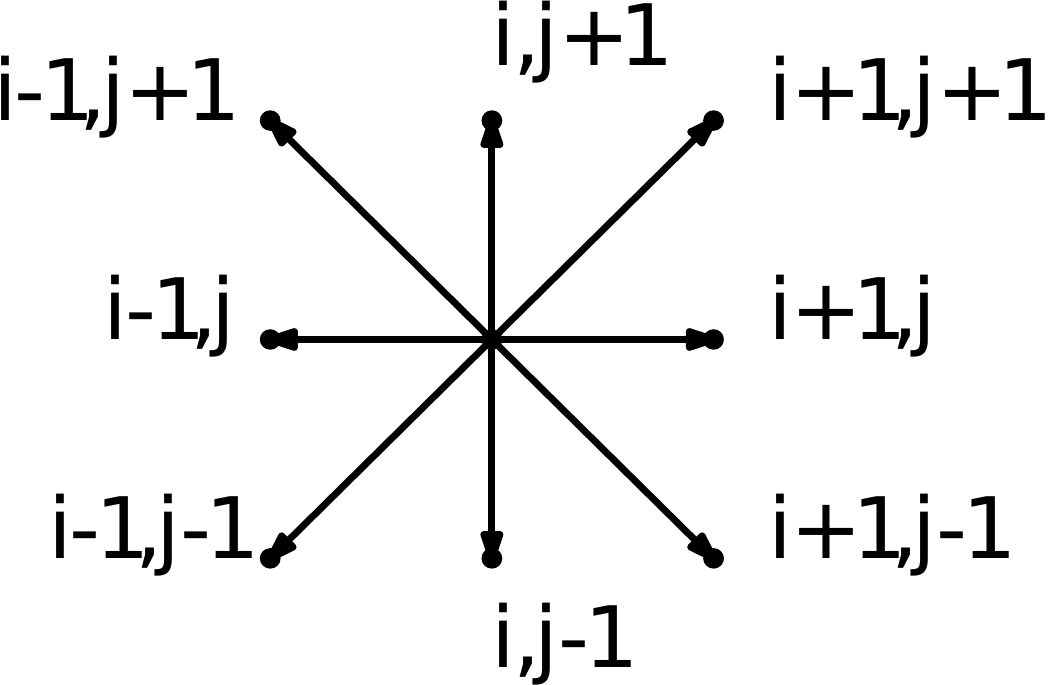} \protect

\caption{\label{lattice} Computation grid corresponding to a discrete gradient.}
\end{figure}

Let us consider a 2-D analog of the formulation depicted in\cite{thampi-iso,rashmi},
where the computational grid is constructed through a sequence of
square lattices. For a 2-D grid with connecting vectors, shown in
Fig.\ref{lattice}, one could explicitly write the discrete gradient
operators as

\begin{equation}
\tilde{\nabla}_{x}\phi=\frac{1}{3\,\delta x}\left(\phi_{i+1,j}-\phi_{i-1,j}\right)+\frac{1}{12\,\delta x}\left(\phi_{i+1,j+1}-\phi_{i-1,j+1}+\phi_{i+1,j-1}-\phi_{i-1,j-1}\right),
\end{equation}
\begin{equation}
\tilde{\nabla}_{y}\phi=\frac{1}{3\,\delta x}\left(\phi_{i,j+1}-\phi_{i,j-1}\right)+\frac{1}{12\,\delta x}\left(\phi_{i+1,j+1}+\phi_{i-1,j+1}-\phi_{i+1,j-1}-\phi_{i-1,j-1}\right).
\end{equation}

Various stencils on which isotropic operators can be written are documented
in \cite{thampi-iso,rashmi}. An implementation of these isotropic
operators in discretization of PDE and SPDE are also used in \cite{thampi,sevink}.

\subsection{Factorizable sign-definite isotropic Laplacians}

With this definition of gradient and Eq.\ref{cd2divgrad} as definition
of Laplacian, we create an FDT preserving isotropic discrete Laplacian
as, $\tilde{\Delta}^{{\rm iso}}=\tilde{\nabla}^{{\rm iso}}\cdot\tilde{\nabla}^{{\rm iso}}$,
which allows us to write an FDT preserving discrete space-time representation
of model B(Eq.\ref{CHC}) as
\begin{equation}
\phi_{i}^{n+1}=\phi_{i}^{n}+\delta t\,M\tilde{\Delta}^{{\rm iso}}(A-K\tilde{\Delta}^{{\rm iso}})\phi_{i}^{n-1}]+\sqrt{\delta t}\tilde{\nabla}^{{\rm iso}}\cdot\boldsymbol{\xi}
.\label{distB}
\end{equation}

\begin{figure}
\centering
\includegraphics[scale=0.5]{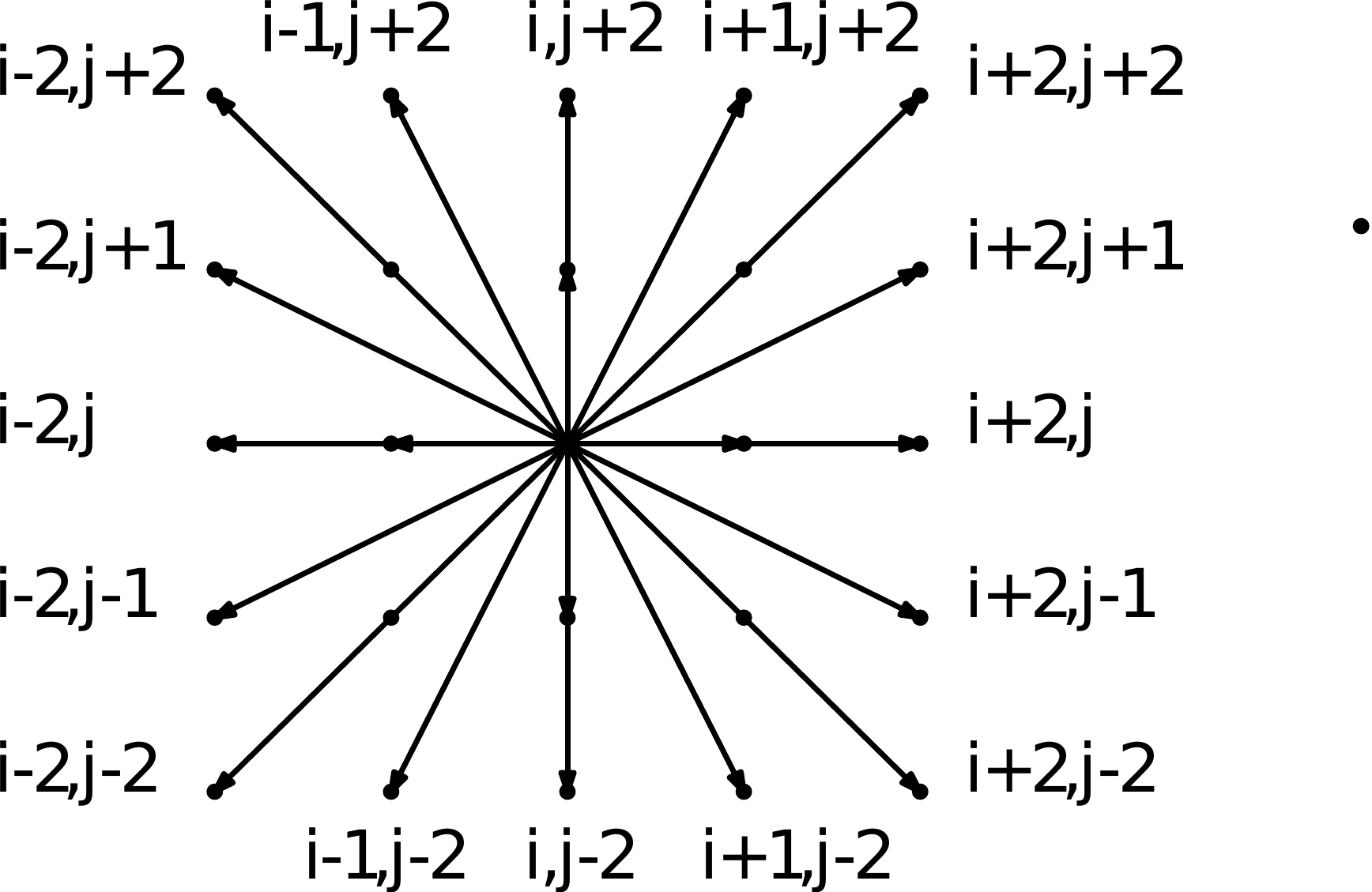}
\caption{\label{latticeLap} Computation grid corresponding to a discrete isotropic
Laplacian.}

\end{figure}

Now, following Eq.\ref{cd2divgrad} one could formulate the discrete
isotropic Laplacian by using $\tilde{\nabla}\cdot\tilde{\nabla}$,
which can be expressed explicitly on a grid, shown in Fig.\ref{latticeLap}
as

\begin{eqnarray}
\tilde{\nabla}^2 & \phi=\frac{\phi_{i+2,j}+\phi_{i-2,j}+\phi_{i,j+2}+\phi_{i,j-2}-4\phi_{i,j}}{9\,\delta x^2}\\ \nonumber
 & -\frac{\phi_{i,j+1}+\phi_{i,j-1}+\phi_{i+1,j}+\phi_{i-1,j}-4\phi_{i,j}}{9\,\delta x^2}\\ \nonumber
 & +\frac{\phi_{i+2,j+2}+\phi_{i-2,j+2}+\phi_{i+2,j-2}+\phi_{i-2,j-2}-4\phi_{i,j}}{72\,\delta x^2} \\ \nonumber
 & +\frac{1}{18\,\delta x^2}\left(\phi_{i+2,j+1}+\phi_{i-2,j+1}+\phi_{i+2,j-1}+\phi_{i-2,j-1}\right.\\ \nonumber
 & \left.+\phi_{i+1,j+2}+\phi_{i+1,j-2}+\phi_{i-1,j+2}+\phi_{i-1,j-2}-8\phi_{i,j}\right),
\end{eqnarray}
which in Fourier domain is a positive quantity
\begin{equation}
\tilde{\nabla}^2(k_{x},k_{y})=-\frac{1}{9}\left[\sin^2(k_{x}\delta_{x})\left(\cos(k_{y}\delta_{y})+2\right)^2+\left(\cos(k_{x}\delta_{x})+2\right)^2\sin^2(k_{y}\delta_{y})\right].
\end{equation}

\section{Model B: Harmonic fluctuations}

For the sake of simplicity and without loss of generality, we first
consider the following simplified case of model B, describing the
dynamics of a non-interacting order parameter in single phase equilibrium,
\begin{equation}
\phi_{i}^{n+1}=\phi_{i}^{n}+\delta t\,D\,\tilde{\Delta}^{{\rm iso}}\phi_{i}^{n-1}+\sqrt{\delta t}\tilde{\nabla}^{{\rm iso}}\cdot\boldsymbol{\xi},\label{stochdiff}
\end{equation}
with $D=A\,M$ being the diffusion coefficient.

\begin{figure}[ht!]
\centering
\includegraphics[scale=0.09]{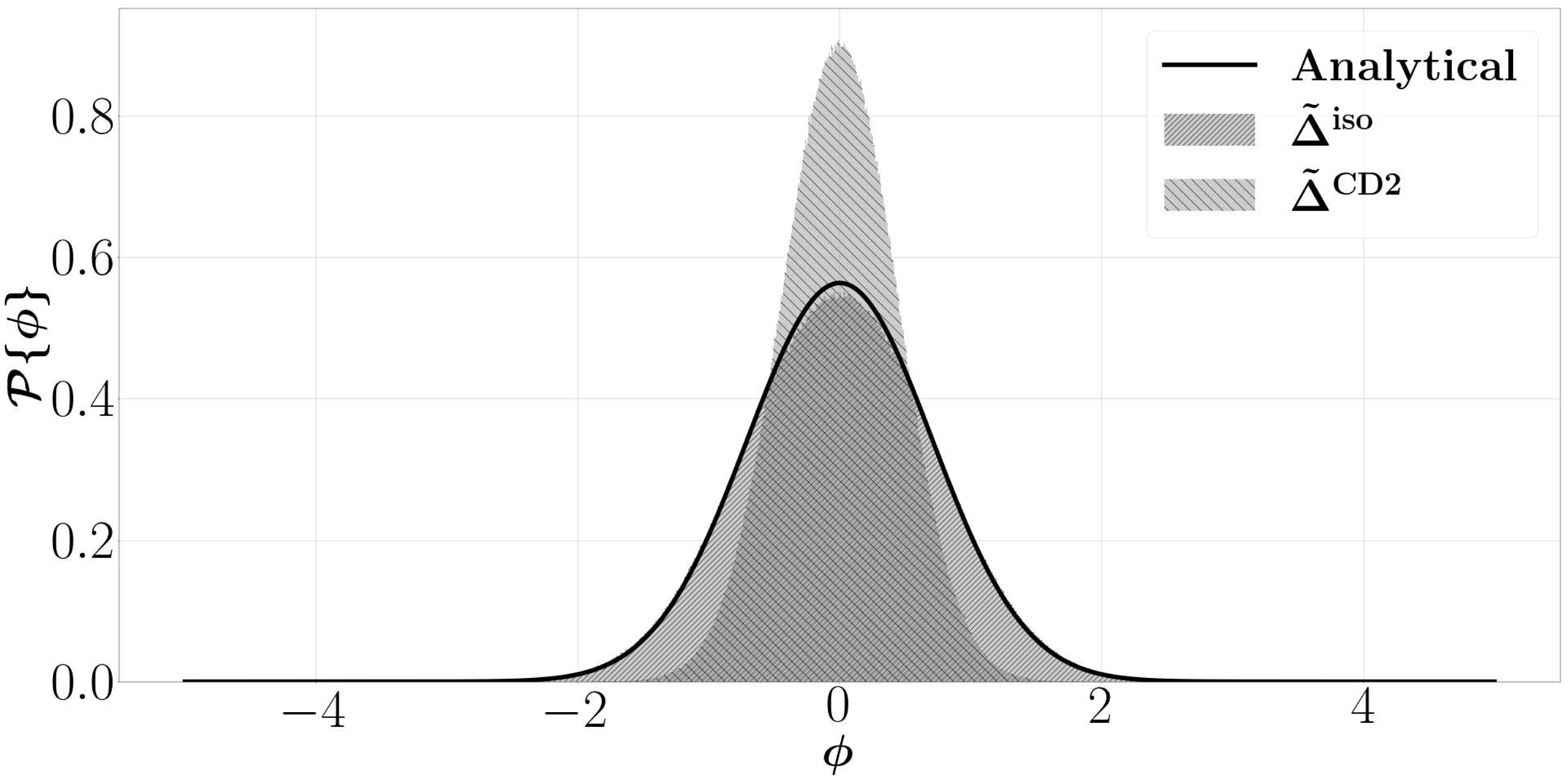} \protect\caption{\label{DistEq} Probability distribution of the order parameter filed
for CD2 ($\alpha=0.05$) and Isotropic schemes ($\alpha=0.22$), computed
by numerically integrating Eq.\ref{CHC} for a grid size of $128\times128$.}
\end{figure}

\begin{figure}[ht!]
\centering
\includegraphics[scale=0.09]{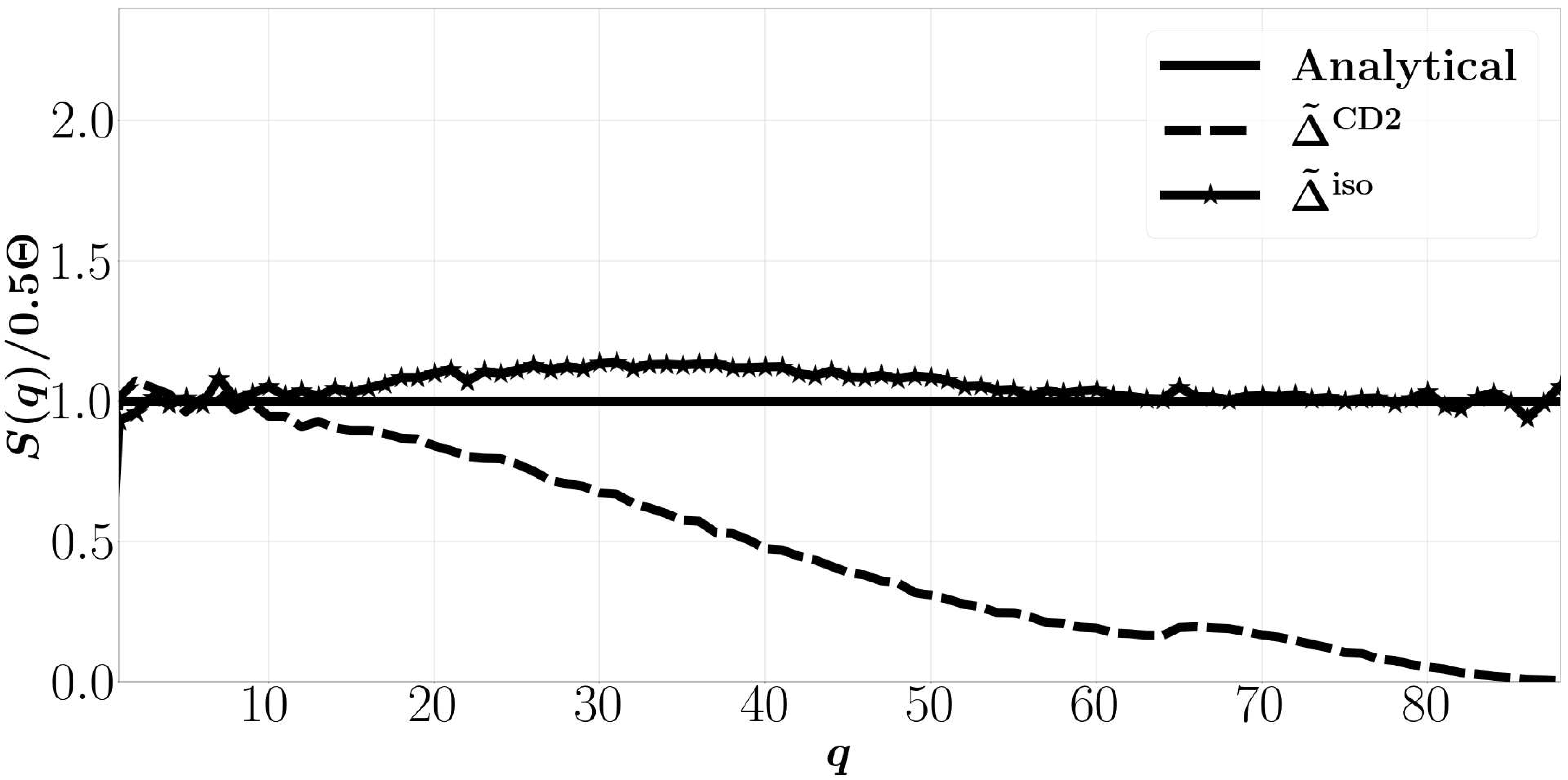} \protect\caption{\label{diffspectra} Spectra of the equilibrium ratio ($S(\mathbf{k})/(0.5k_{B}T))$)
for CD2 ($\alpha=0.05$) and Isotropic schemes ($\alpha=0.22$), computed
by numerically integrating Eq.\ref{CHC} for a grid size of $128\times128$.}
\end{figure}

We contrast the present approach with the traditional discretization
schemes, by performing a long time integration for this setup
where the steady state probability distribution can be compared with
the Gaussian distribution expected from the continuous model B dynamics.
The superiority of the present work is apparent in the Fig.\ref{DistEq},
where the the probability distribution of the order parameter $\phi(\mathbf{x},t)$
is plotted. It should be noted that, with a considerably larger($\alpha=0.22$)
time step than that of the conventional CD2, current scheme shows
much better agreement with the Gaussian. We also compare the spectra
of the normalized energy distribution(equilibrium ratio) in Fig.\ref{diffspectra}
for the CD2 and present scheme with the analytical one, from which
the break down of FDT at discrete level for CD2 is quite apparent.

\begin{figure}[h!]
\centering
\includegraphics[scale=0.08]{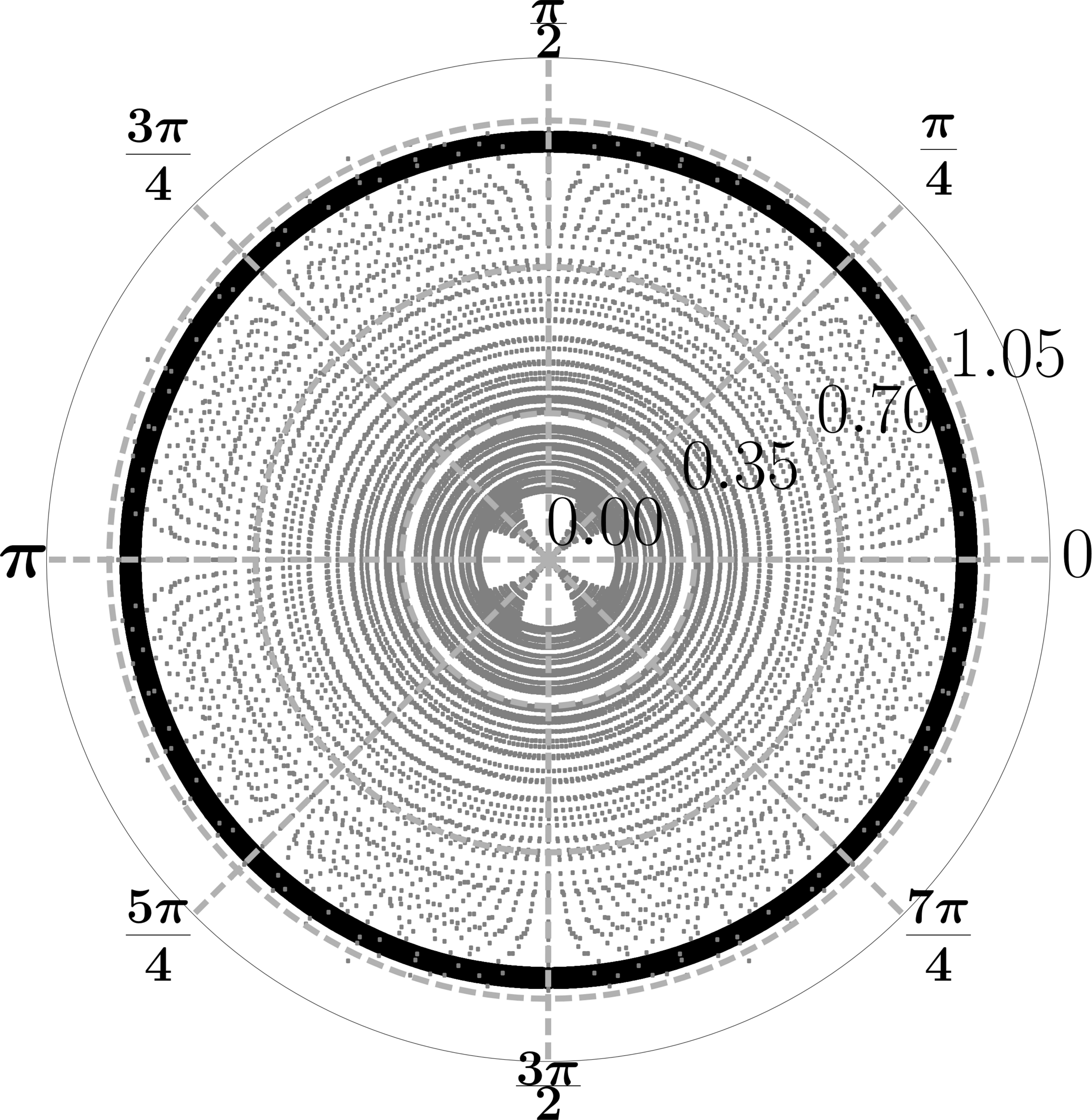} 
\includegraphics[scale=0.08]{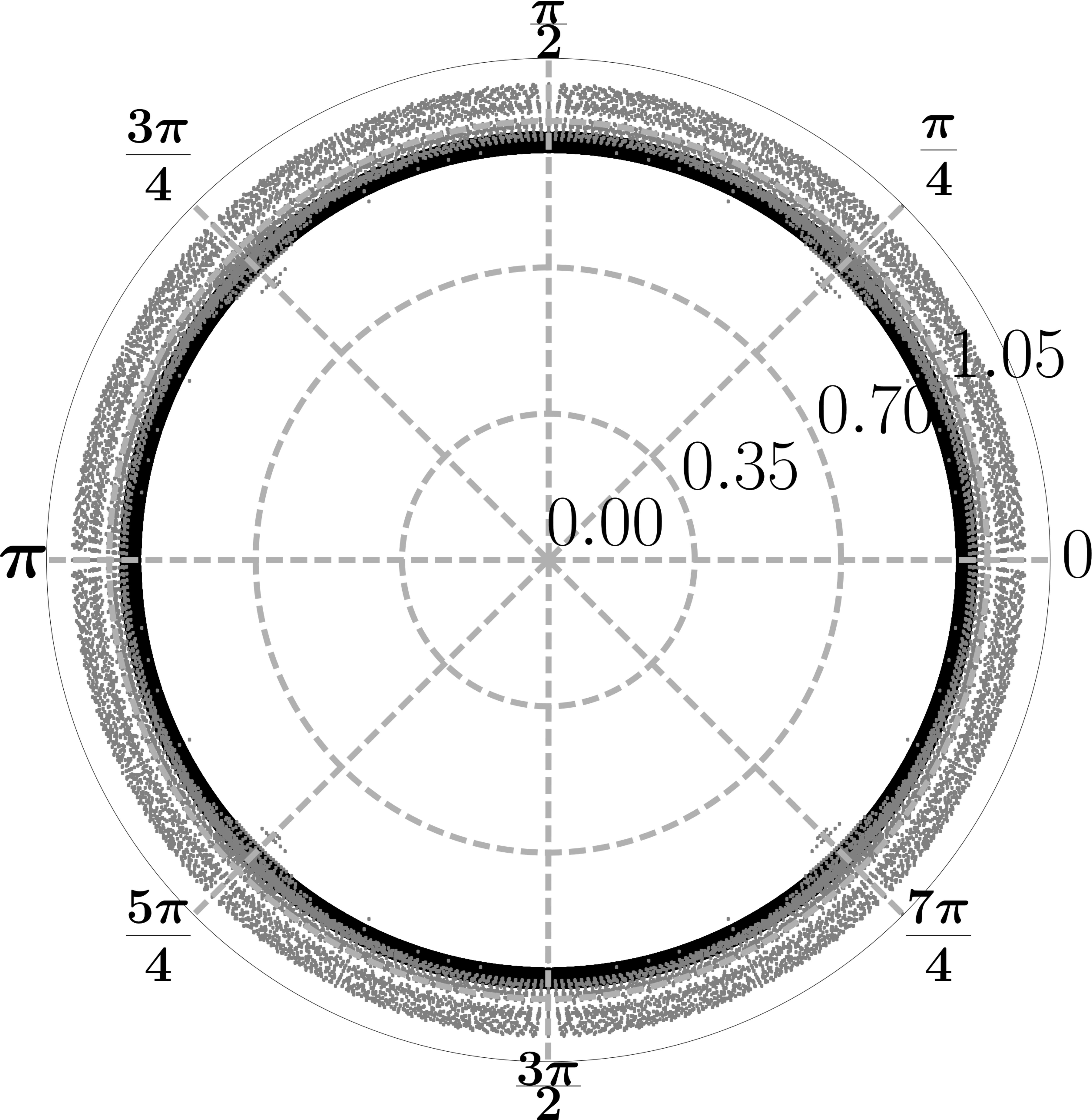} \protect\caption{\label{polarEn} Polar plot of the equilibrium ratio ($S(\mathbf{k})/(0.5k_{B}T))$)
at different wave numbers. Left: $\tilde{\Delta}^{{\rm CD2}}$ with
$\alpha=0.05$, Right: $\tilde{\Delta}^{{\rm iso}}$ with $\alpha=0.22$,
Black circle: Analytical, computed by numerically integrating Eq.\ref{CHC}
for a grid size of $128\times128$.}
\end{figure}

To highlight the quantitative improvements due to the present work,
we present the polar plots of the normalized energy distribution(equilibrium
ratio) at different wave numbers in Fig.\ref{polarEn}. We remind
that for model B with non-interacting single well Landau-Ginzburg
Hamiltonian of the order parameter, energy in any wave number is $0.5k_{B}T$
irrespective of the wave number, thus the normalized energy at any
wave number should be one. As expected, result from current scheme
is quite close to the analytical result, while central discretization
shows violation of FDT even with the use of a time step which is four
times smaller than that of the present scheme.

\section{Model B: Anharmonic fluctuations}

In this section, we extend our treatment to the case of a inhomogeneous
system, where different regions are coupled via free-energy gradients
of the order parameter entering in the Landau-Ginzburg Hamiltonian
${\cal F}$. Use of the aforementioned isotropic
operators and delayed discretization, the discrete model B assumes
the following form, with $f(\phi_{i}^{n})$ denoting the value of
the respective discrete free energy density at lattice cite $i$ and
time step $n$.
\begin{equation}
\phi_{i}^{n+1}=\phi_{i}^{n}+\delta t\,M\tilde{\Delta}^{{\rm iso}}\left[f(\phi_{i}^{n-1})-K\tilde{\Delta}^{{\rm iso}}\phi_{i}^{n-1}\right]\label{replidel}+\sqrt{\delta t}\tilde{\nabla}^{{\rm iso}}\cdot\boldsymbol{\xi}.\nonumber
\end{equation}

\begin{figure}
\centering
\includegraphics[scale=0.09]{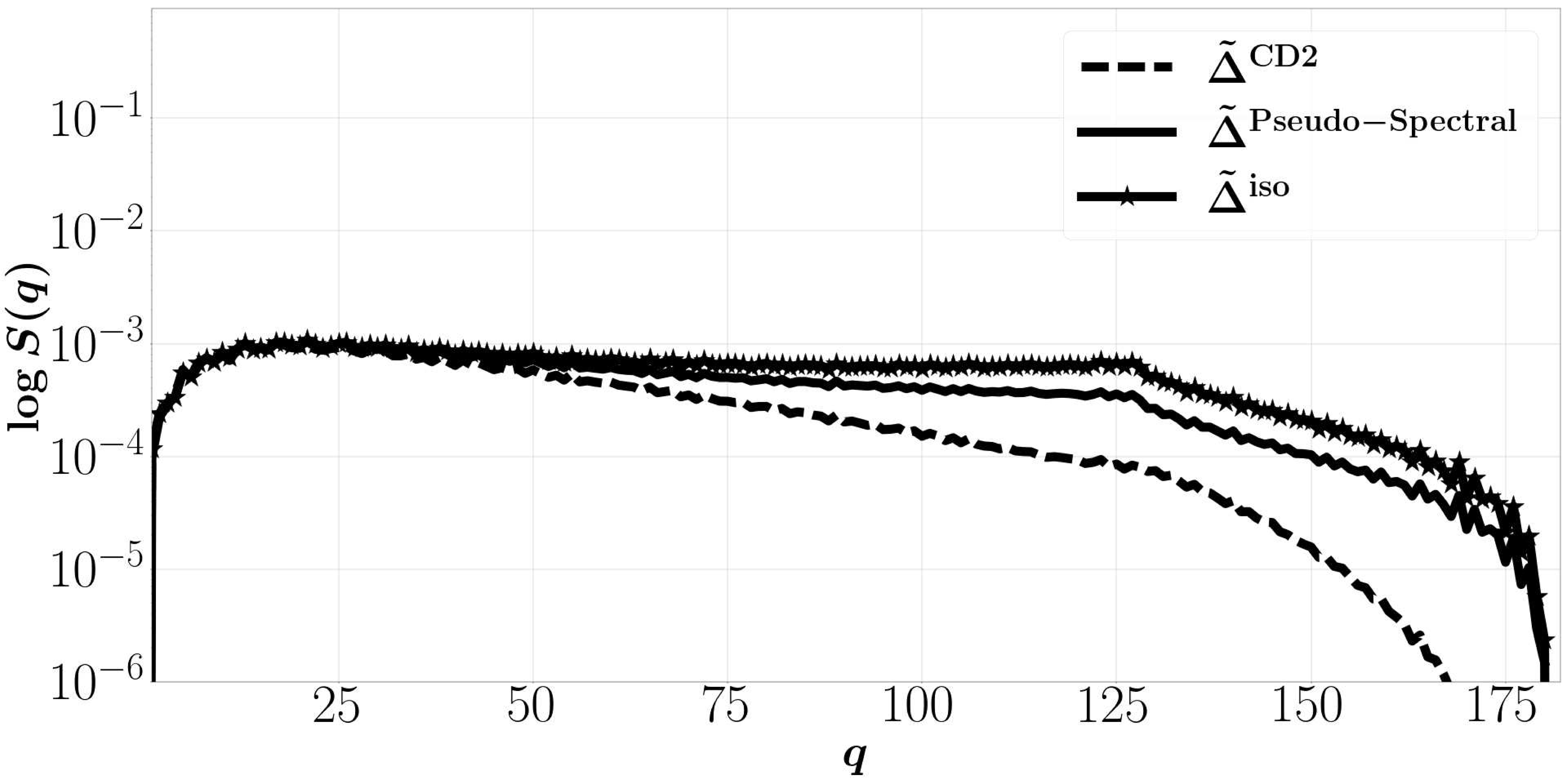} \quad{}\includegraphics[scale=0.09]{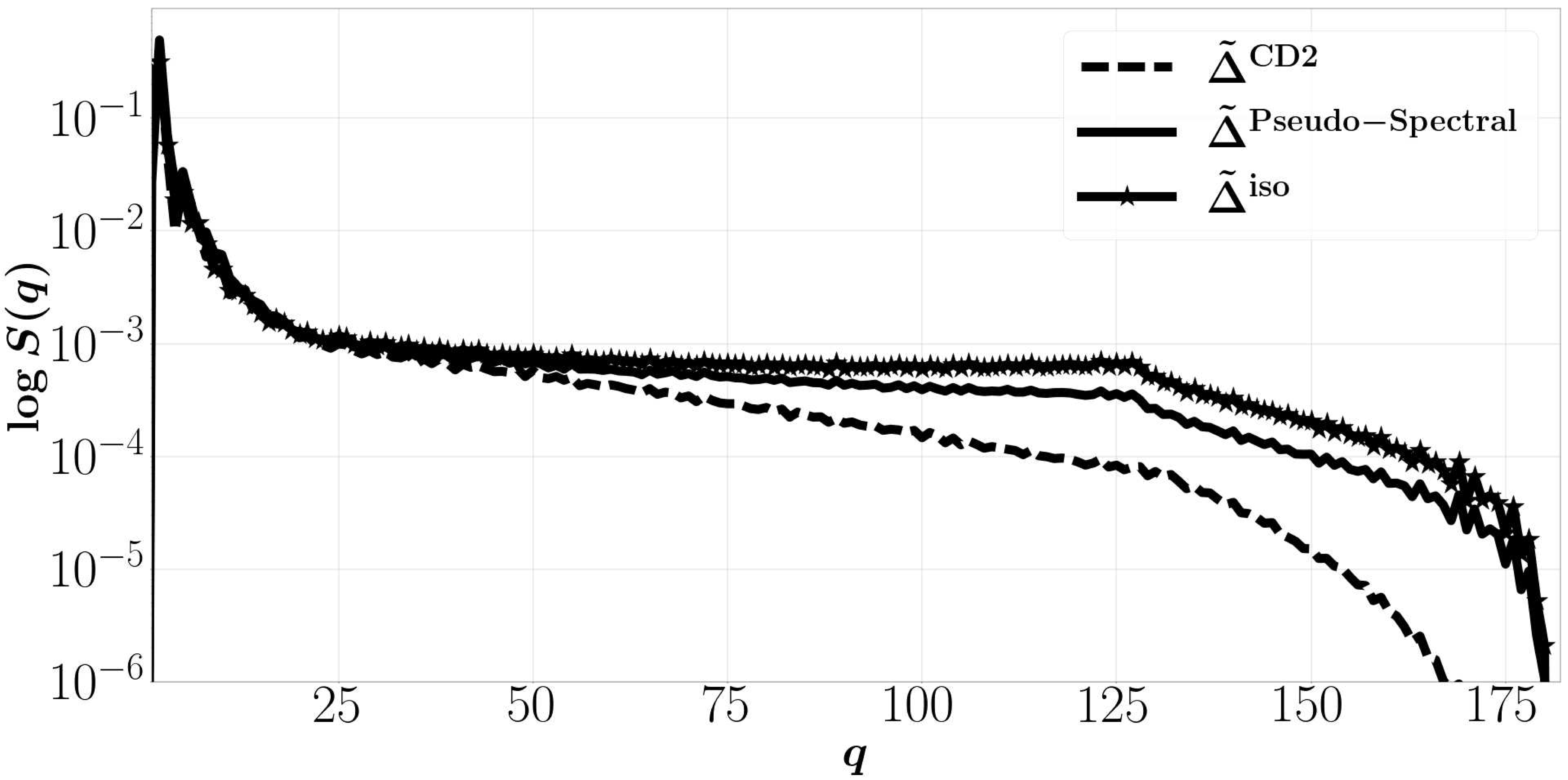}
\protect\caption{\label{chc_spectra} Time averaged spectra of the single well(top)
and double well(bottom) structure factor $S(\mathbf{k})$ for, CD2($\alpha=0.01$),
Isotropic($\alpha=0.04$) and Pseudo-Spectral($\alpha=0.005$), computed
by numerically integrating Eq.\ref{replidel} for a grid size of $256\times256$.}
\end{figure}

\begin{figure}
\centering
\includegraphics[scale=0.06]{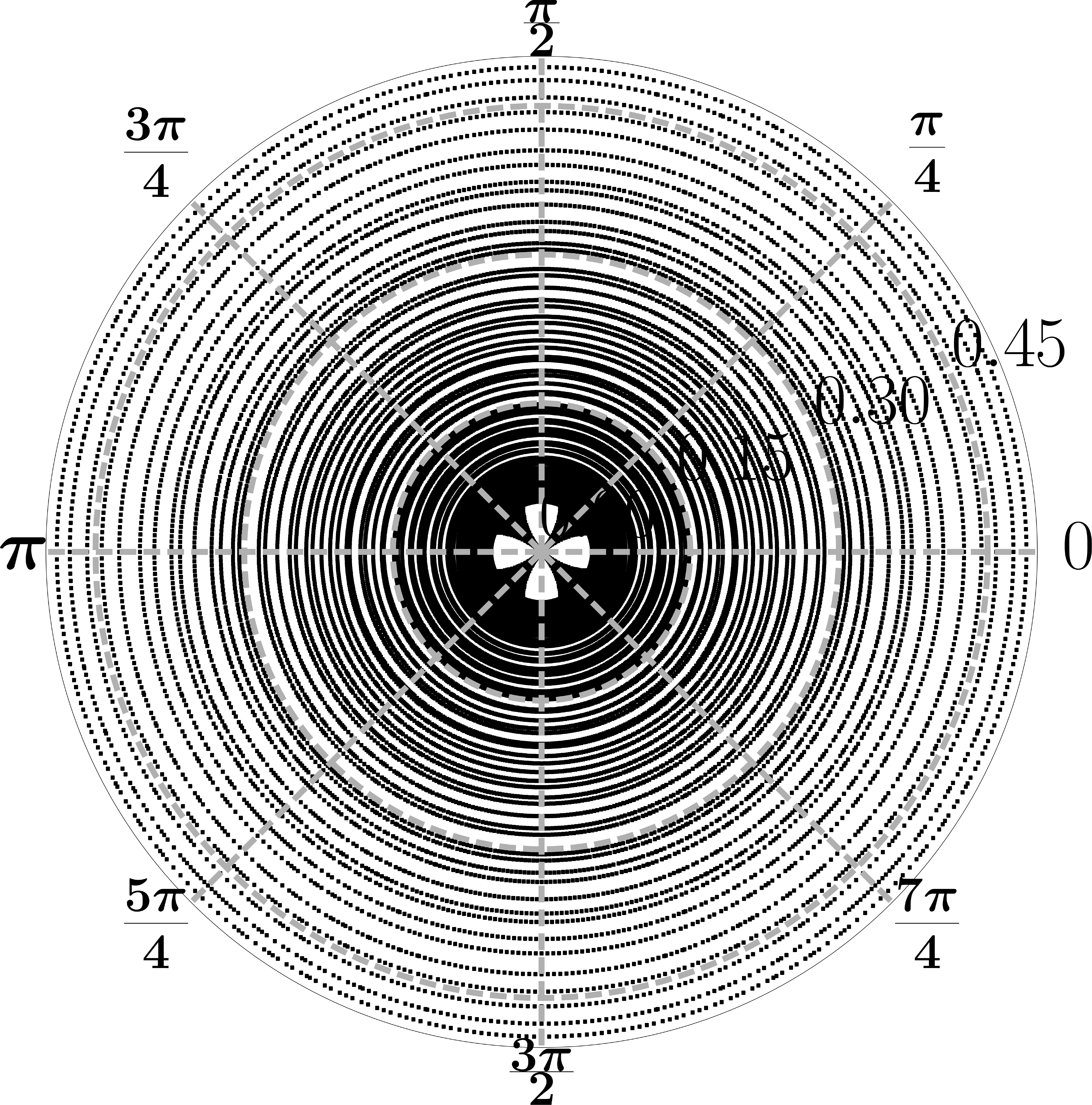} \quad \includegraphics[scale=0.06]{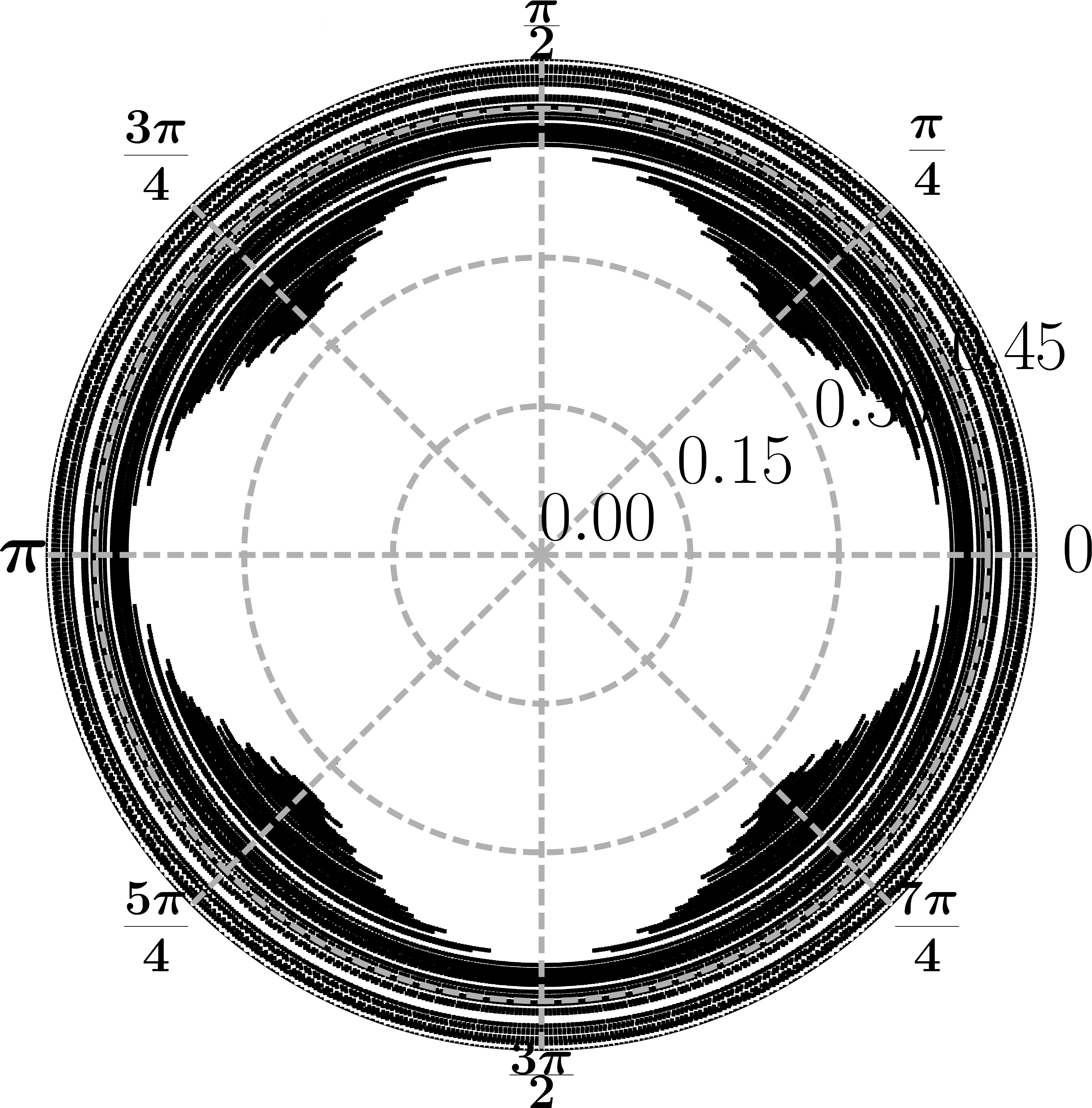}
\quad
\includegraphics[scale=0.06]{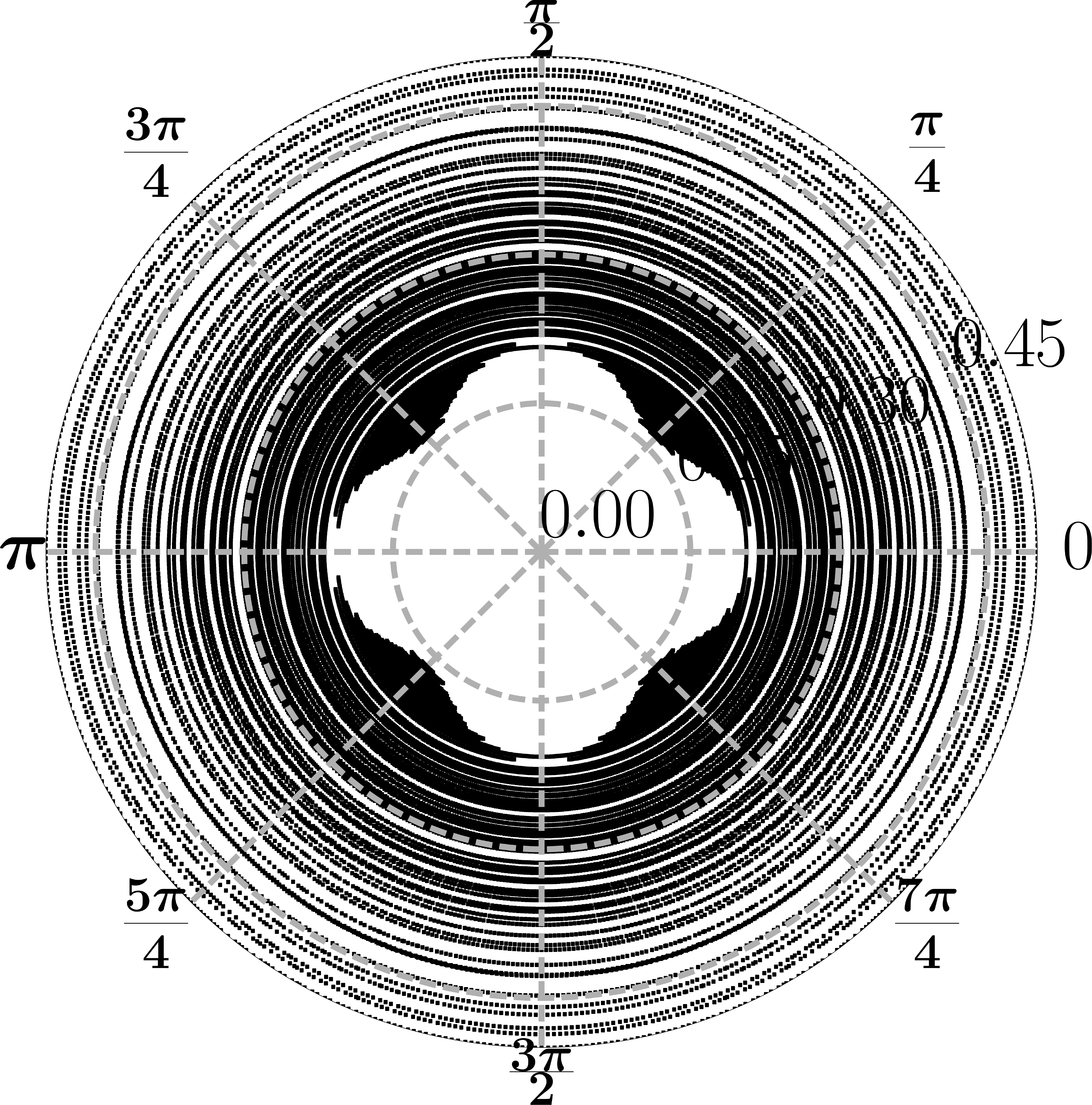} \protect\caption{\label{spolarEn} Polar plot of the single well structure factor $S(\mathbf{k})$
at different wave numbers. Left: $\tilde{\Delta}^{{\rm CD2}}$($\alpha=0.01$),
Middle: $\tilde{\Delta}^{{\rm iso}}$($\alpha=0.04$), Right: Pseudo-Spectral($\alpha=0.005$),
computed by numerically integrating Eq.\ref{replidel} for a grid
size of $256\times256$. }
\end{figure}

\begin{figure}
\centering
\includegraphics[scale=0.06]{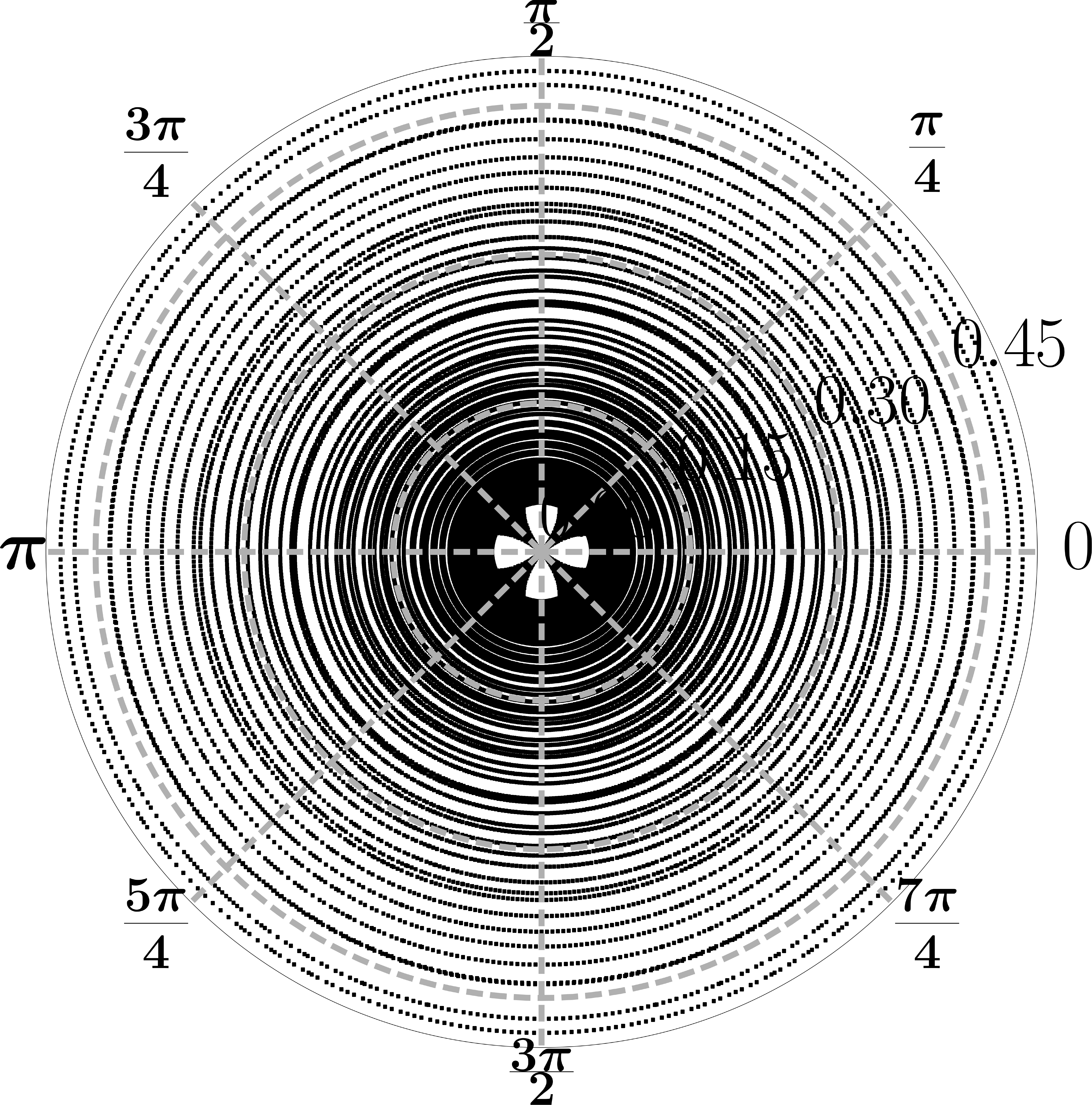} \quad \includegraphics[scale=0.06]{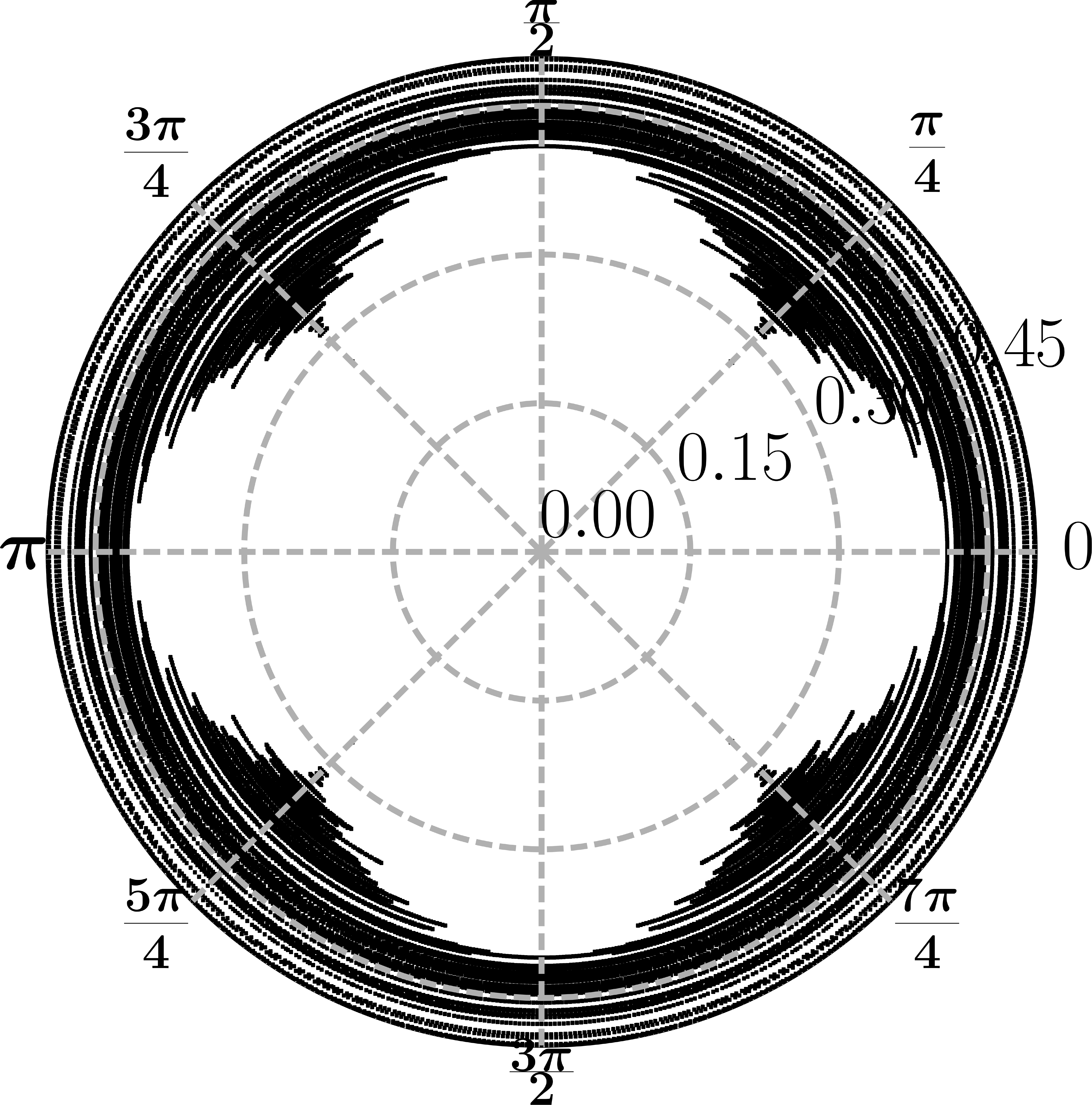}
\quad
\includegraphics[scale=0.06]{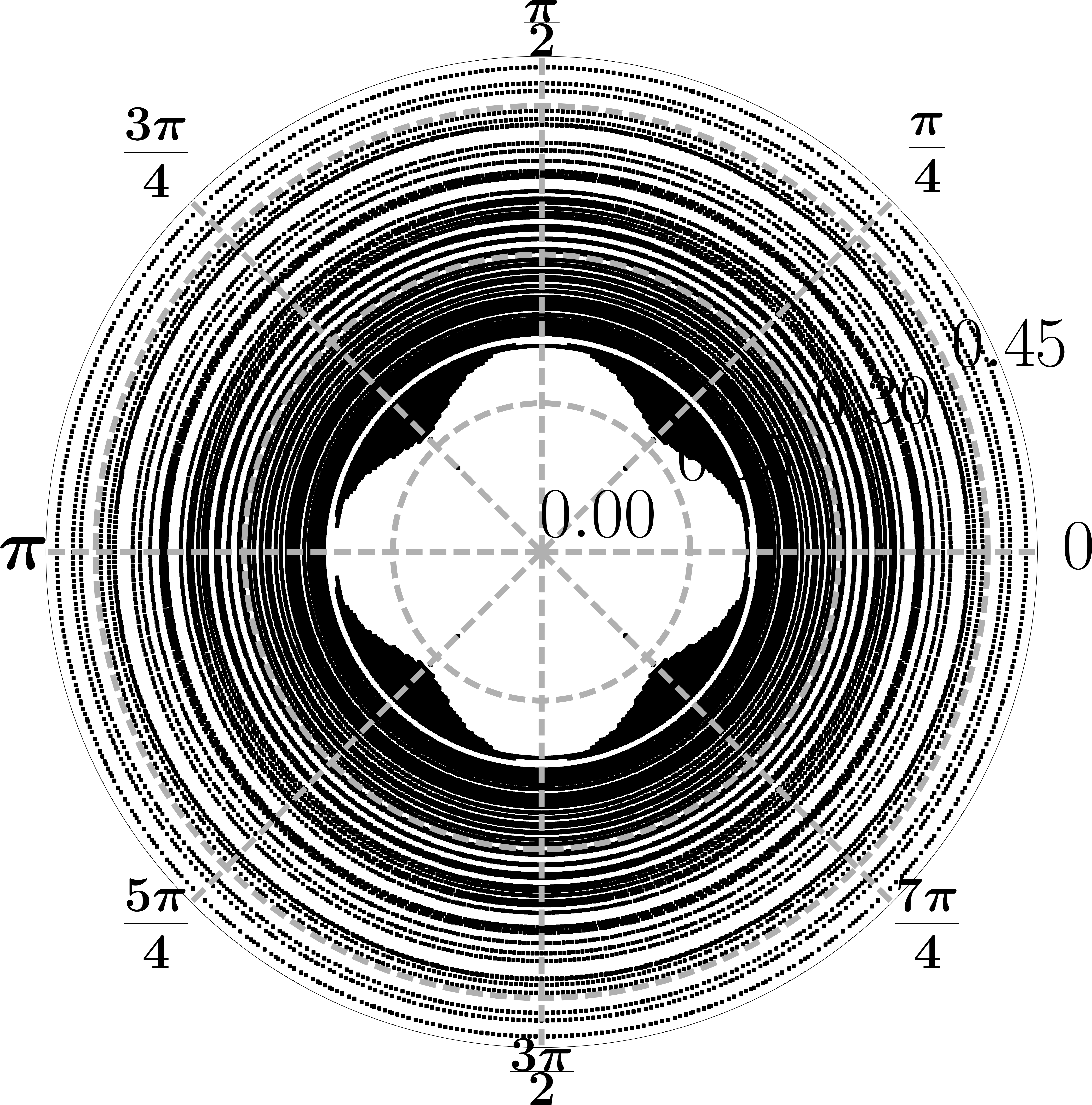} \protect\caption{\label{dpolarEn} Polar plot of the double well structure factor $S(\mathbf{k})$
at different wave numbers. Left: $\tilde{\Delta}^{{\rm CD2}}$($\alpha=0.01$),
Middle: $\tilde{\Delta}^{{\rm iso}}$($\alpha=0.04$), Right: Pseudo-Spectral($\alpha=0.005$),
computed by numerically integrating Eq.\ref{replidel} for a grid
size of $256\times256$, computed by numerically integrating Eq.\ref{replidel}
for a grid size of $256\times256$. }
\end{figure}

To establish the consistency of the present work over the usual central
difference type operators, we compare the spectra of the structure
factors obtained from three different discrete formulation of model
B, viz. isotropic, central difference and Fourier pseudo spectral
in Fig.\ref{chc_spectra}. Here free energy densities for both single
phase $f(\phi)=\frac{1}{2}A\phi^2$ and two phase equilibrium $f(\phi)=\frac{1}{2}A\phi^2+\frac{1}{4}B\phi^{4}$
are considered. The pseudo spectral, ensuring exact space derivatives, preserve the FDT at discrete level but the computation is much more expensive than the other two counterparts viz. CD2 and isotropic. On the other hand it is evident here that the failure of preserving
FDT at the lattice level leads to an energy loss at the higher wave
numbers for the case of CD2 as compared to that of the pseudo spectral result.
The isotropic formulation does not show any such energy
loss at the higher wave numbers, instead its energy spectra is very close to that of the pseudo spectral. To
illustrate the quantitative aspect of the present formulation over
the traditional schemes we also present the polar spectra in Fig.\ref{spolarEn}
and Fig.\ref{dpolarEn} for the structure factors of these three cases,
which clearly bring out the anisotropy and breakdown of FDT at the
discrete level for the traditional CD2 schemes as opposed to the isotropic
discretization.

It should be stressed here that in all these cases the isotropic scheme
operates at a time step four times larger than that of the CD2 scheme.

\begin{figure}
\begin{center}
 \includegraphics[scale=0.15]{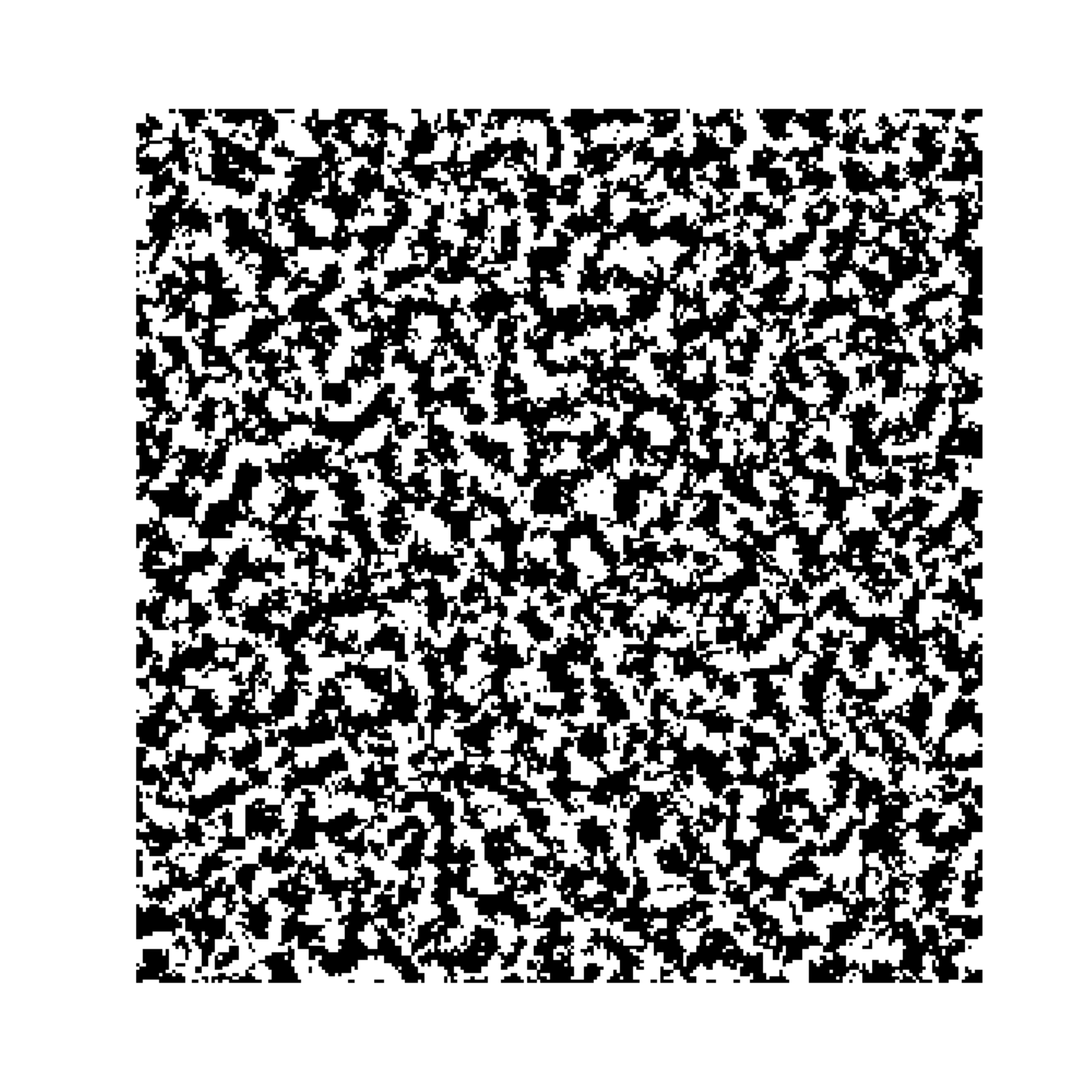}\includegraphics[scale=0.15]{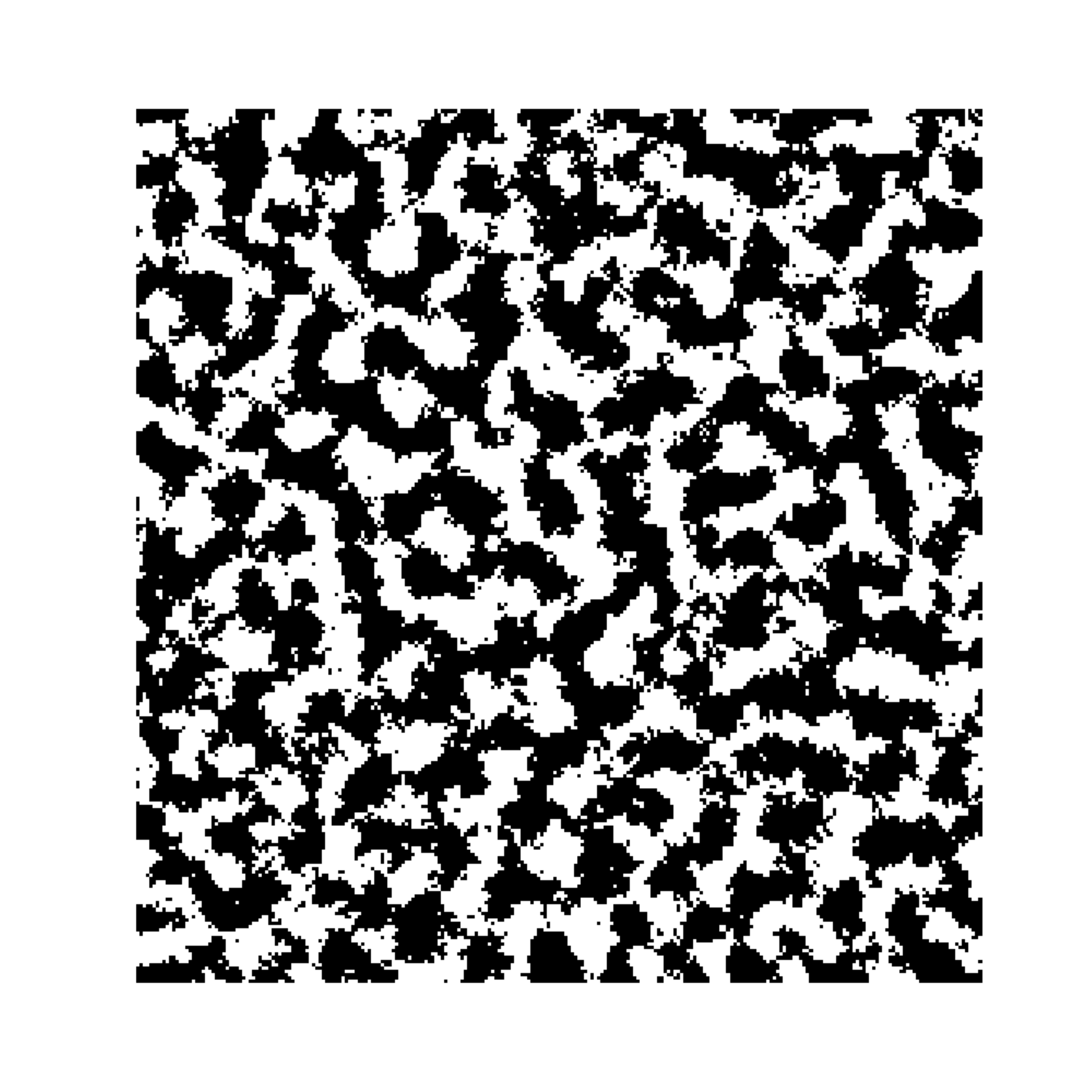}
\includegraphics[scale=0.15]{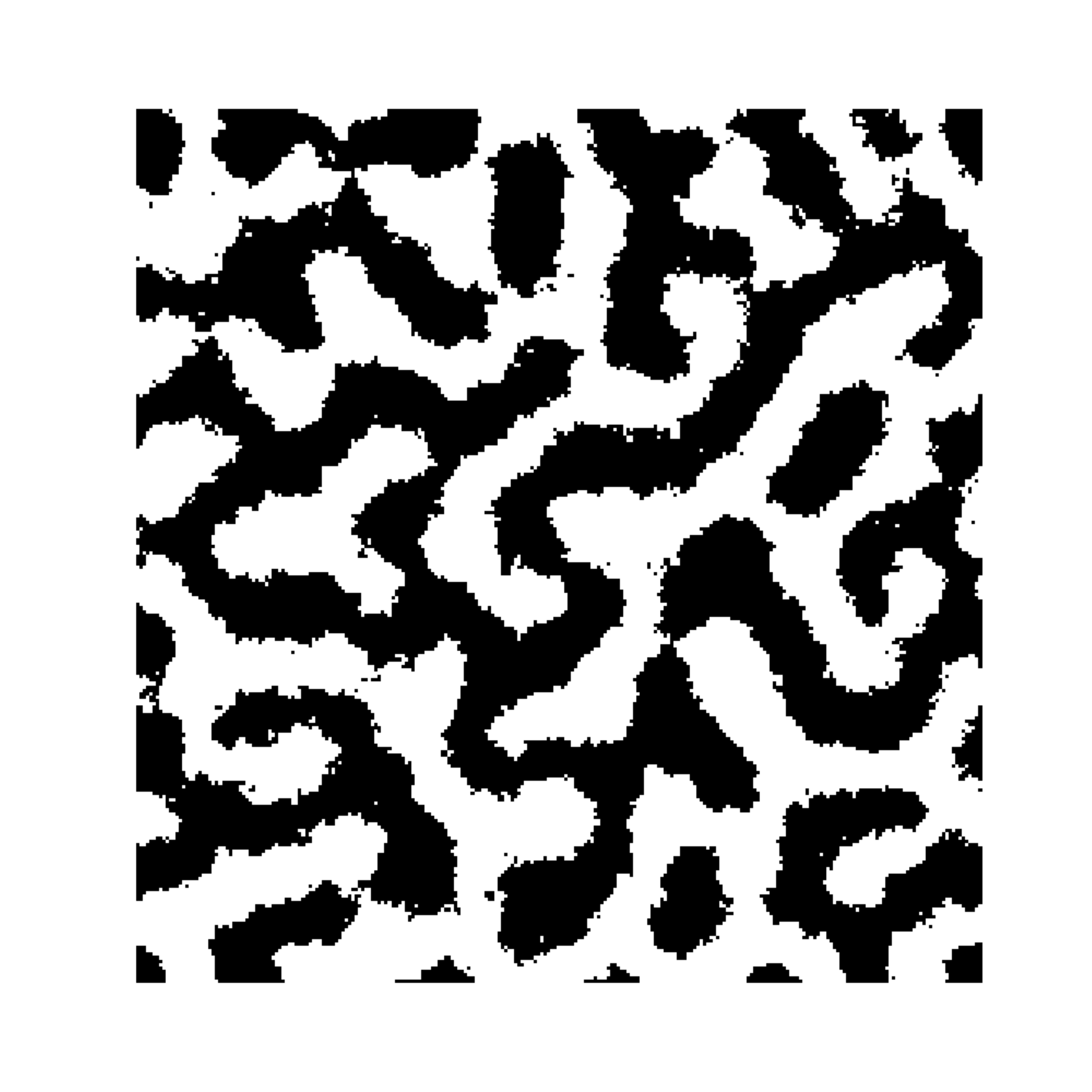} \\
 \includegraphics[scale=0.15]{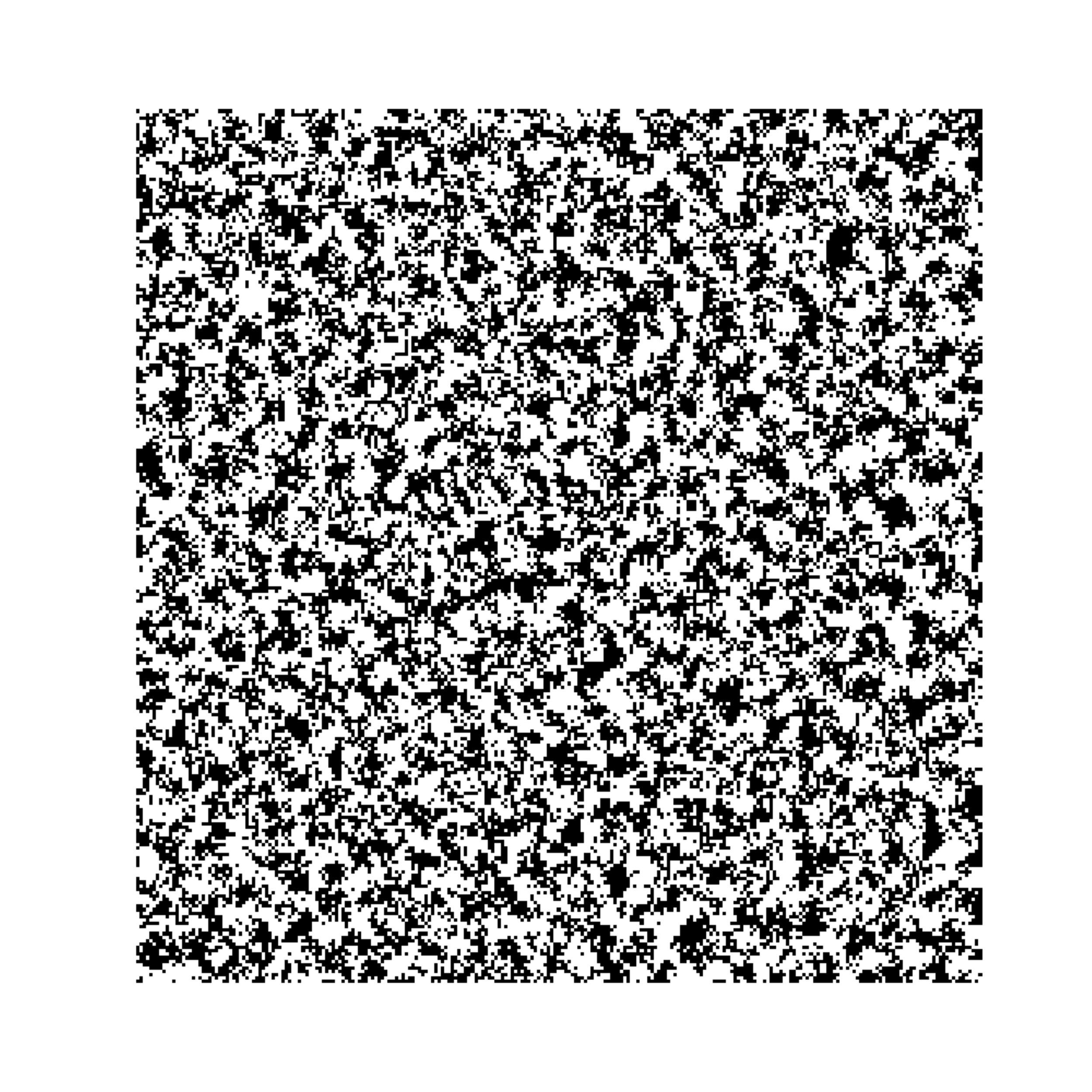}\includegraphics[scale=0.15]{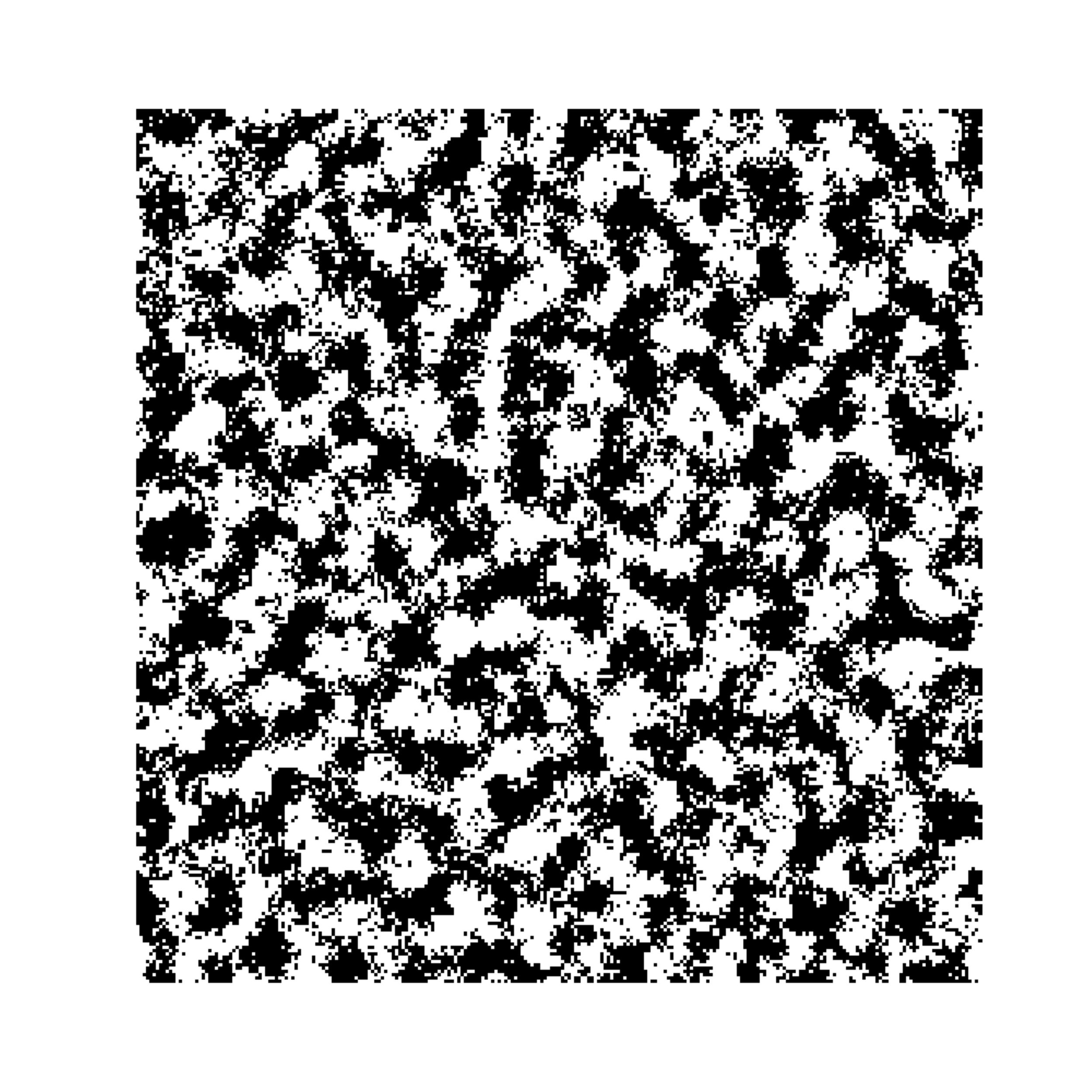}
\includegraphics[scale=0.15]{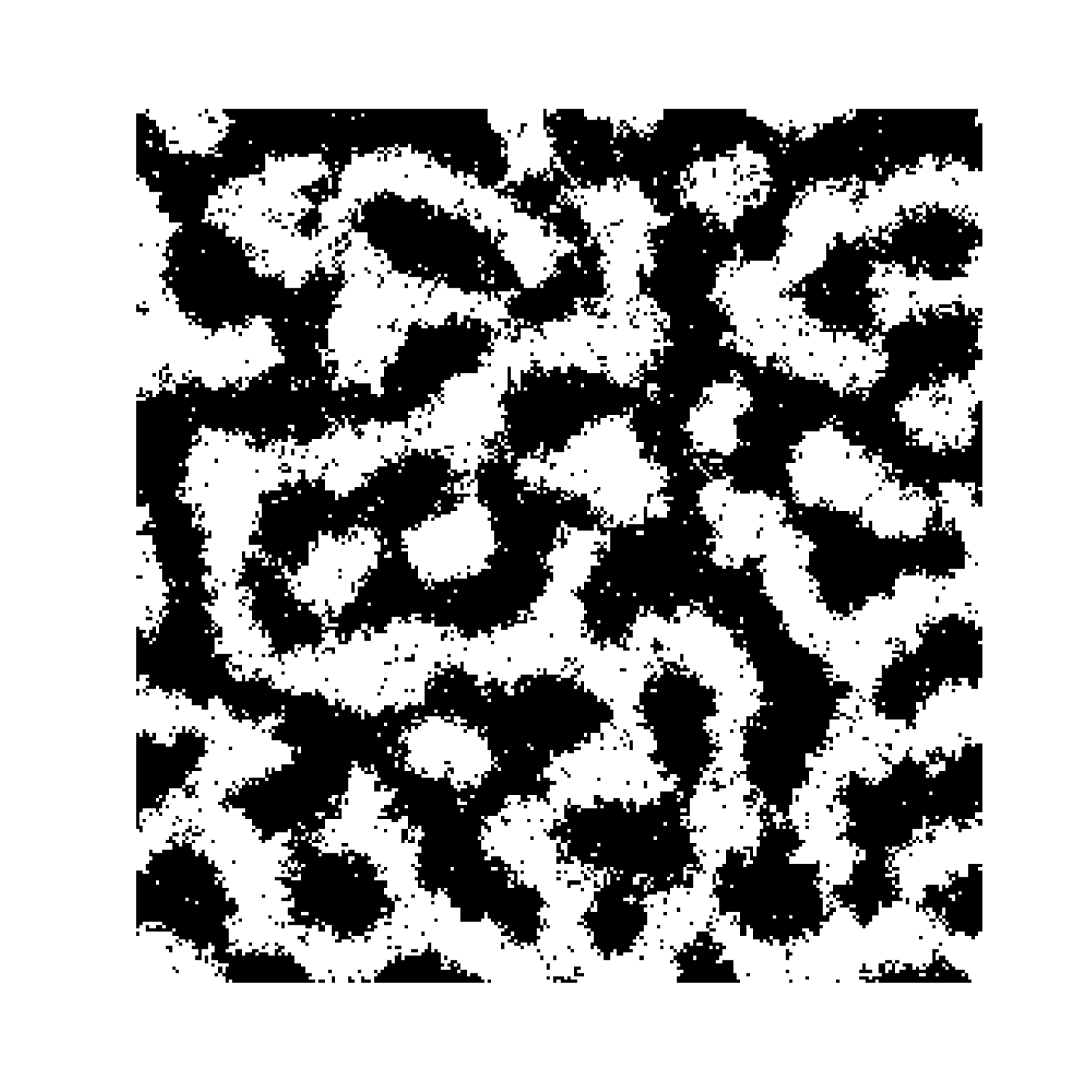} \\
 \includegraphics[scale=0.15]{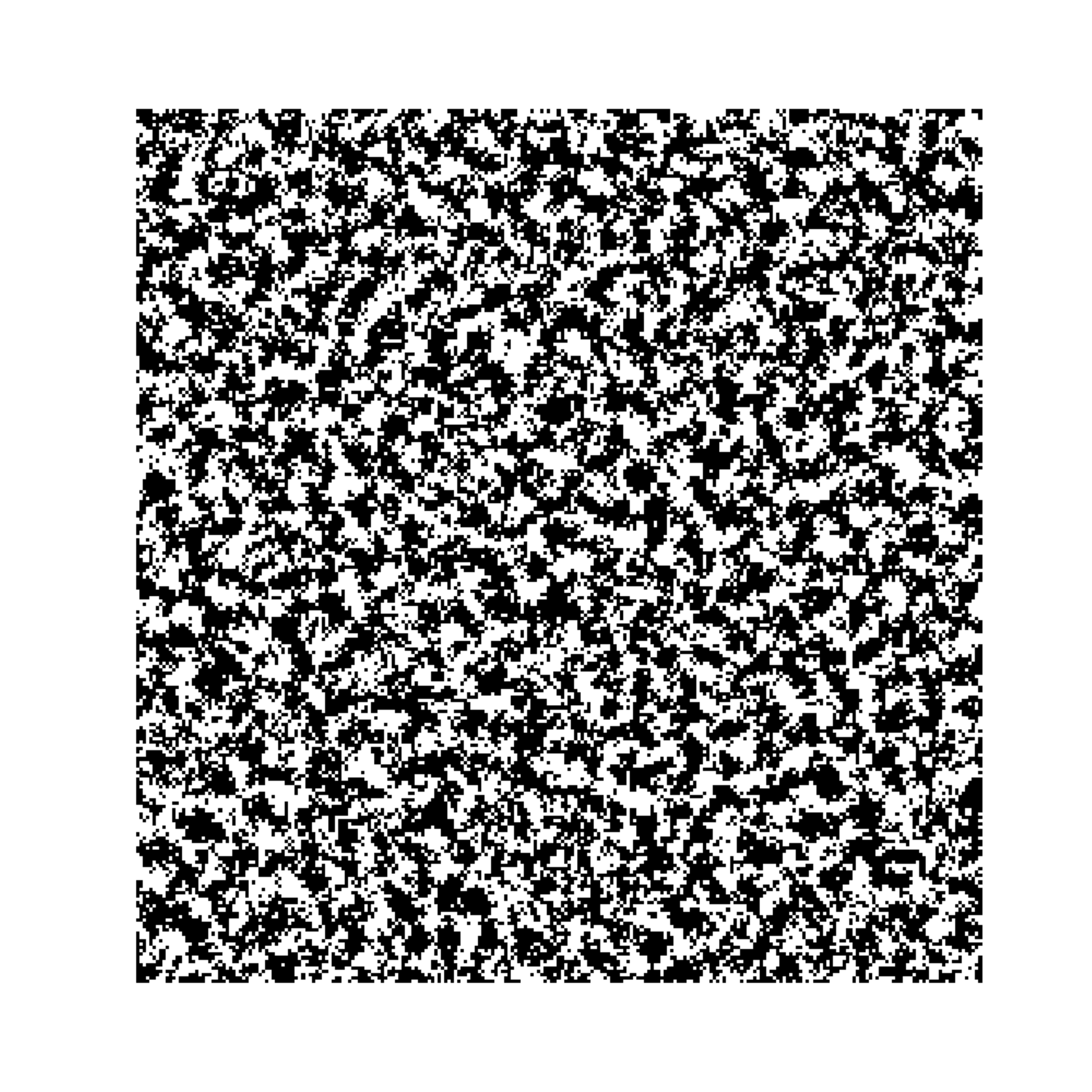}\includegraphics[scale=0.15]{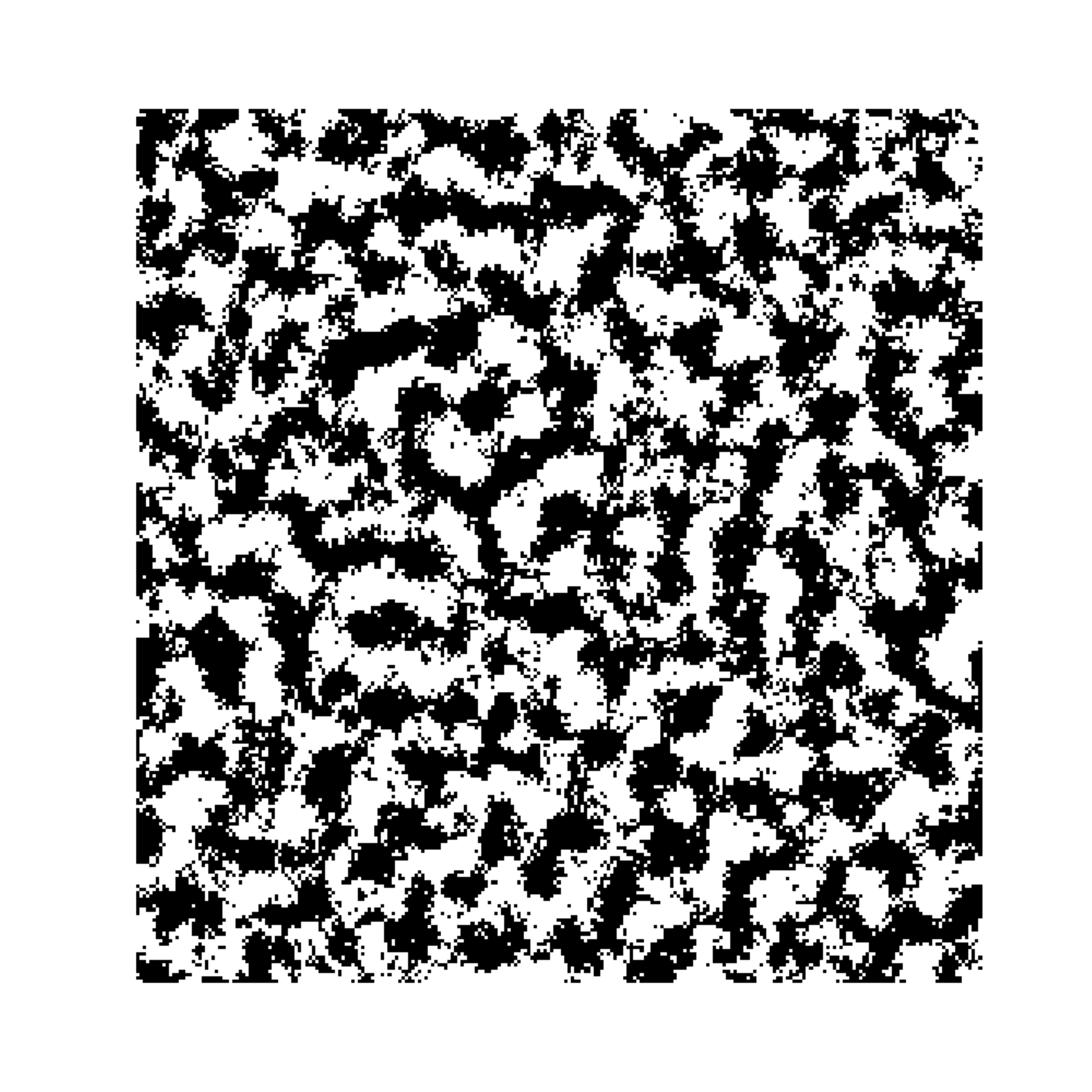}
\includegraphics[scale=0.15]{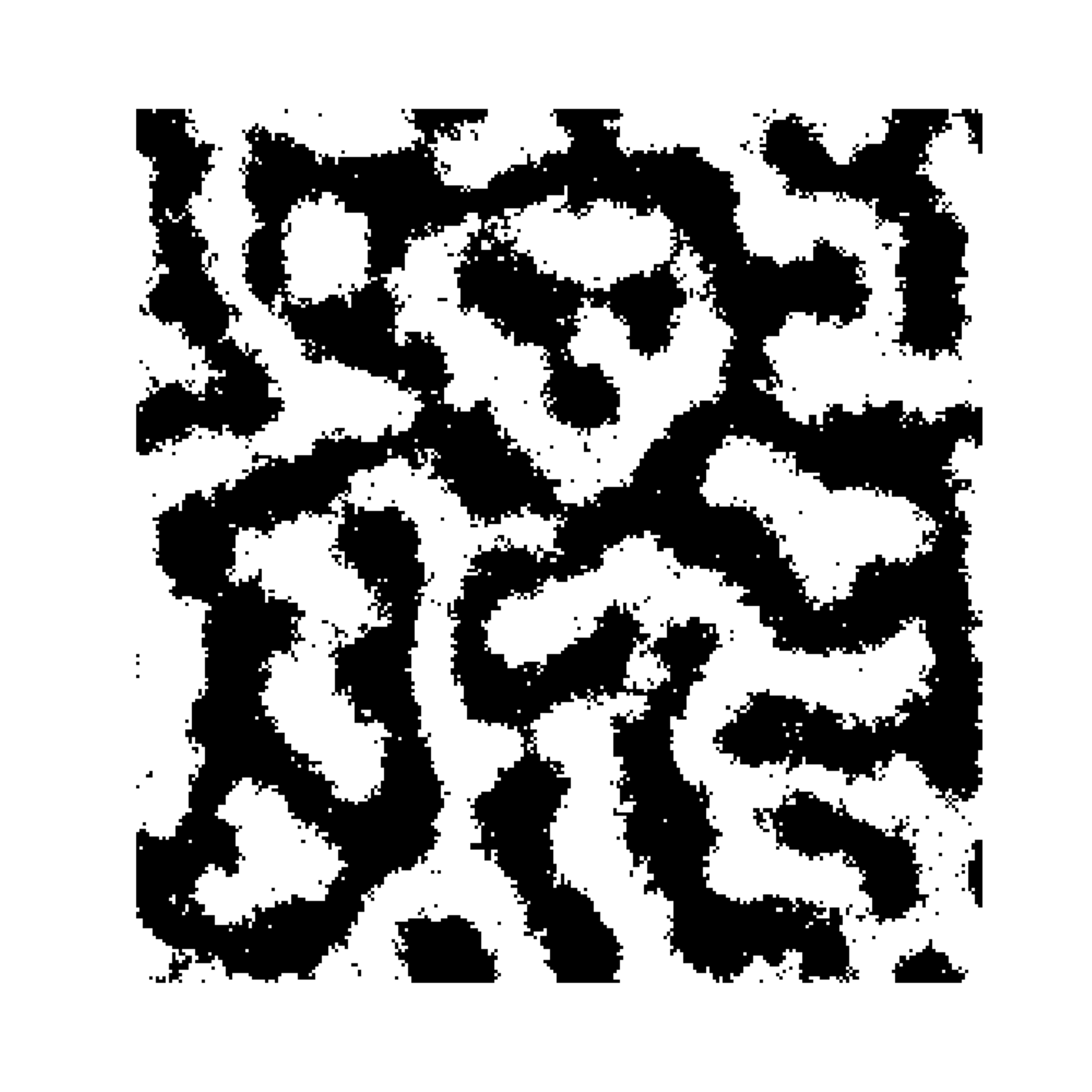} \protect\caption{\label{phase} Order parameter dynamics at three different instances
for CD2(top panel), Isotropic and Pseudo-Spectral(bottom panel). Each panel shows three different instances, viz. $t=10,100,1000$. Here, CD2
seems to be more quenched than the other two.}
\end{center}
\end{figure}

To characterize the diffusive behaviour of the central difference
scheme over the present formulation, we investigate the dynamics of
the order parameter $\phi$ through its instantaneous distribution.
In Fig.\ref{phase} three consecutive instantaneous states of the order
parameter evolution are compared for three different cases, namely
CD2, isotropic and pseudo-spectral.

It is evident from this plot that, for CD2, the system is in a little
more quenched state than that of the isotropic or pseudo-spectral
methods. This is indicative of the fact that the CD2 has more diffusive
behaviour at higher wave numbers, which is also observed in Fig.\ref{chc_spectra}.

\section{Outlook}

To conclude, we have presented a discrete framework where the essence
of phase separation dynamics in terms of fluctuation-dissipation relation
is preserved. Thus, similar to the cell dynamical method, a fully
self-consistent framework at discrete level is obtained. The present
approach also allows the discrete framework to inherit transport properties
and free energies known from PDE based formulation.

It should also be pointed out that, the present formulation can easily
be extended to three dimensions(3D), as all of these aforementioned
isotropic operators can also be derived in 3D. For a detailed formulation
of these operators in higher dimensions one might refer to \cite{thampi,rashmi}.
Also in order to contrast the present formulation with cell dynamics,
we write Eq.\ref{distB} in a form analogous to Eq.(2.7,2.8) of
\cite{oono1988} as:
\begin{equation}
\phi_{i}^{n+1}= \phi_{i}^{n}+D\,\delta t\left[\left\langle \left\langle \phi_{i}^{n-1}\right\rangle \right\rangle -\phi_{i}^{n-1}\right]+\sqrt{dt}\tilde{\nabla}^{{\rm iso}}\cdot\pmb\xi
,\label{CDS}
\end{equation}
where,
\begin{equation}
\tilde{\nabla}^{{\rm iso}}\phi(\mathbf{x},t)=\frac{1}{3\delta x}\sum_{i=1}^{N_{i}}\phi+\frac{1}{12\delta x}\sum_{j=1}^{N_{j}}\phi
,\label{CDSgrad}
\end{equation}
\begin{equation}
\left\langle \left\langle \phi(\mathbf{x},t)\right\rangle \right\rangle = -\frac{1}{9}\sum_{i=1}^{N_{i}}\phi+\frac{1}{9}\sum_{j=1}^{N_{j}}\phi+\frac{1}{18}\sum_{k=1}^{N_{k}}\phi+\frac{1}{72}\sum_{l=1}^{N_{l}}\phi+\frac{1}{2}\phi(\mathbf{x},t).\label{CDSlap}
\end{equation}
Here, $N_{i},N_{j},N_{k},N_{l}$ are the nearest and next nearest
neighbours and so on. It can be observed that the discrete form of
the model B in Eq.\ref{CDS} and the discrete operators Eq.\ref{CDSgrad}
and Eq.\ref{CDSlap} preserve the same structure of a Cell Dynamical
System(CDS)\cite{oono1988}. Though the form is similar, two key differences,
wider stencil and use of past data are must be noted. These two differences
allow us to formulate a cell dynamical system where FDT is preserved
even at the discrete level and the connection with the PDE is also
apparent.

\section*{Reference}
\bibliographystyle{iopart-num}
\bibliography{lattice-fdt}


\end{document}